\documentclass[aps,prd,twocolumn,showpacs,superscriptaddress,nofootinbib,preprintnumbers]{revtex4-2}

\usepackage{amssymb,amsmath,verbatim,mathtools,needspace,enumitem,etoolbox,graphicx,physics,microtype,afterpage,xspace,tabularx,lmodern,multirow}
\usepackage{booktabs}
\usepackage{siunitx}
\usepackage{gensymb}
\usepackage{ulem}  

\usepackage{amsmath}
\usepackage{tikz}
\usepackage{tikz-feynman}
\usepackage{booktabs}
\usepackage{bm}
\usepackage{cancel}
\tikzfeynmanset{
  compat=1.1.0
}
\tikzset{every path/.append style={thick}}
\usepackage{amsfonts}
\usepackage{amssymb}
\usepackage{color}
\usepackage{graphicx}
\usepackage{bm}
\usepackage[colorlinks=true,citecolor=blue,urlcolor=blue]{hyperref}
\hypersetup{colorlinks=true,citecolor=romared,linkcolor=romared,urlcolor=romared}
\definecolor{romared}{RGB}{142,0,28}

\usepackage{afterpage}
\usepackage{multirow}
\usepackage{appendix}
\usepackage[table]{xcolor}
\usepackage{feynmp-auto}
\usepackage{aas_macros}
\usepackage{makecell}
\usepackage{soul}

\usepackage{lipsum}
\usepackage{orcidlink}
\definecolor{tabblue}{RGB}{31, 119, 180}
\definecolor{darkblue}{RGB}{0, 0, 120}
\definecolor{tabred}{RGB}{214, 39, 40}
\definecolor{tabgreen}{RGB}{44, 160, 44}
\definecolor{tabgray}{RGB}{100, 100, 100}
\definecolor{taborange}{RGB}{255, 127, 14}
\definecolor{tabbrown}{RGB}{128, 0, 0}
\definecolor{tabpink}{RGB}{255, 141, 161}
\definecolor{tabpurple}{RGB}{148, 103, 189}
\definecolor{goldenrod}{RGB}{218, 165, 32}
\usepackage{float}
\usepackage{placeins}
\begin{document}

\preprint{\hbox{UTWI-10-2026}}

\title{Parameterizing Dark Energy at the density level: A two-parameter alternative to CPL}
\author{Gabriele Montefalcone\,\orcidlink{0000-0002-6794-9064}}
\email{montefalcone@utexas.edu}
\affiliation{Texas Center for Cosmology and Astroparticle Physics, Weinberg Institute for Theoretical Physics, \\
Department of Physics, The University of Texas at Austin, Austin, TX 78712, USA}

\author{Richard Stiskalek\,\orcidlink{0000-0002-0986-314X}}
\email{richard.stiskalek@physics.ox.ac.uk}
\affiliation{Astrophysics, University of Oxford, Denys Wilkinson Building, Keble Road, Oxford, OX1 3RH, UK}

\begin{abstract}
\noindent
We introduce a minimal two-parameter formulation of the dark energy (DE) density evolution normalized to its present-day value, $f_{\rm DE}(z) \equiv \rho_{\rm DE}(z)/\rho_{\rm DE,0}$, in terms of $f_p\equiv f_{\rm DE}(z_p)$ and the DE equation of state $w_p\equiv w(z_p)$, at a pivot redshift $z_p$. This provides an alternative framework for assessing the evidence for evolving DE, complementary to the established Chevallier-Polarski-Linder (CPL) parameterization of the DE equation of state in terms of $w_0$ and $w_a$. By parameterizing the DE density directly, the $(w_p,\,f_p)$ formulation avoids the approximate degeneracies intrinsic to the $(w_0,\,w_a)$ basis---in particular the weak sensitivity of the expansion history to $w_a$---while reproducing the background evolution of representative quintessence models with equivalent accuracy.
Confronting it with the latest baryon acoustic oscillation (BAO) measurements from DESI, a prior on early-universe parameters from {\it Planck}
cosmic microwave background (CMB) observations, and Type Ia supernovae (SNe) data, we find that the $w_p$ and $f_p$ parameters are both tightly constrained and sensitive to distinct subsets of the data. Specifically, $w_p$ is measured to percent-level precision by BAO and CMB alone, while $f_p$ is pinned down by the independent matter density constraint that only SNe provide. Including the Pantheon\texttt{+} SNe sample, we obtain $w_p = -1.04 \pm 0.04$ and $f_p = 1.07 \pm 0.04$, with similar results when using the DESY5 SNe sample. The preference for evolving DE over $\Lambda$CDM remains below $3\sigma$ across all dataset combinations, comparable to that obtained with CPL. Notably, the proximity of both $w_p$ and $f_p$ to their cosmological constant values of $(-1,1)$---precisely at the epoch where the data are most sensitive---deepens the coincidence previously identified in the CPL framework, reinforcing the case for caution in interpreting the current evidence for dynamical DE.
\end{abstract}

\maketitle

\section{Introduction}

\noindent Recent measurements from the Dark Energy Spectroscopic Instrument (DESI)~\cite{DESI:2024mwx,DESI:2025zgx} have reignited the possibility that cosmic acceleration is driven by a dynamical dark energy (DE) component rather than a cosmological constant.
When combined with cosmic microwave background (CMB) and Type Ia supernovae (SNe) observations, the DESI DR2 baryon acoustic oscillation (BAO) data yield a ${\sim}\,3\sigma$ preference for evolving DE~\cite{DESI:2025zgx, DES:2025tir,Hoyt:2026fve}, assuming spatial flatness and the Chevallier--Polarski--Linder (CPL) parameterization of the DE equation of state~\cite{Chevallier:2000qy,Linder:2002et},
\begin{equation}
    w_{\rm CPL}(z)=w_0+w_a\frac{z}{1+z}\,. \label{eq:cpl}
\end{equation}
The inferred evolution favors a phantom crossing, with the effective equation of state transitioning from $w < -1$ at earlier times to $w > -1$ before the present epoch~\cite{DESI:2025fii}.

These findings have prompted extensive investigation into the nature and robustness of the apparent preference for dynamical DE, employing alternative DE parameterizations~\cite{Tada:2024znt,Luongo:2024fww,Cortes:2024lgw,Carloni:2024zpl,Wang:2024dka,Yang:2024kdo,Chan-GyungPark:2024mlx,Shlivko:2024llw,Dinda:2024kjf,DESI:2024kob,Bhattacharya:2024hep,Ramadan:2024kmn,Roy:2024kni,Gialamas:2024lyw,Malekjani:2024bgi,Najafi:2024qzm,Ye:2024ywg,Giare:2024gpk,Jiang:2024xnu,Reboucas:2024smm,RoyChoudhury:2024wri,Giare:2024oil,Park:2024pew,Li:2024qus,Wolf:2025jlc,Shajib:2025tpd,Giare:2025pzu,Paliathanasis:2025cuc,DESI:2025fii,Shlivko:2025fgv,Efstathiou:2025tie,Khoury:2025txd,Ormondroyd:2025iaf,Paliathanasis:2025dcr,Pang:2025lvh,Chaussidon:2025npr,Kessler:2025kju,Santos:2025wiv,DESI:2025wyn,Wolf:2025jed,Li:2025cxn,Teixeira:2025czm,Paliathanasis:2025hjw,RoyChoudhury:2025dhe,Specogna:2025guo,Scherer:2025esj,Liu:2025mub,Wang:2025vfb,Cheng:2025lod,Cheng:2025hug,Sabogal:2025jbo,Cai:2025mas,Li:2025ula,Ozulker:2025ehg,Li:2025ops,Li:2025dwz,Lee:2025pzo,Silva:2025twg,Ishak:2025cay,Wu:2025vfs,Zhou:2025nkb,Fazzari:2025lzd,Smith:2025icl,Li:2025vuh,Zhang:2025lam,Cheng:2025yue,Ginat:2026fpo,Ladeira:2026jne,Luciano:2026vhm,Paliathanasis:2026ymi,Li:2026xaz, Gokcen:2026pkq, Najafi:2026kxs}, varying the choice of priors on the DE parameters~\cite{Steinhardt:2025znn,Bayat:2025xfr, Toomey:2025xyo, Toomey:2025yuy, Shlivko:2025krk} or the statistical framework under which the evidence is assessed~\cite{Herold:2025hkb,Ong:2025utx,Ong:2026tta}, and using full-shape galaxy clustering data in place of BAO alone~\cite{DESI:2024hhd,Chudaykin:2025aux, Chudaykin:2025lww, Ivanov:2026dvl, Chudaykin:2026nls}.
While the reconstructed evolution of $w(z)$ remains qualitatively stable across these approaches, the statistical significance of the departure from $\Lambda$ can differ considerably (see e.g.~\cite{Wang:2025vfb,Ong:2025utx,Toomey:2025xyo, Chudaykin:2026nls, Weiner:2026sfm}).
Moreover, as noted in Refs.~\cite{Cortes:2024lgw,Efstathiou:2025tie}, reparameterizing CPL in terms of the equation of state $w_p$ evaluated at the pivot redshift---the epoch where a given dataset combination most tightly constrains DE---reveals that $w_p$ is consistent with $-1$ to percent-level precision, with the entire preference for evolving DE driven by the weakly constrained slope $w_a$.
That these models require a substantially varying $w(z)$ to reproduce the data yet return a cosmological constant value at the redshift where the data are most constraining constitutes a new cosmic coincidence, warranting caution in interpreting the current evidence for evolving DE.

Indeed, this preference for evolving DE is partly shaped by the choice of parameterizing at the level of the DE equation of state, which enters the cosmological distances measured by BAO, CMB, and SNe only through a nonlinear integral mapping to the DE density.
This mapping can amplify parameter uncertainties, introduce approximate degeneracies, and smear out features in the underlying evolution~\cite{Maor:2000jy,Barger:2000jg,Wang:2004ru,Shlivko:2024llw}.
The recent analysis of Ref.~\cite{Wang:2025vfb} corroborates this concern: bypassing the equation-of-state parameterization entirely and reconstructing the DE density evolution directly from DESI and CMB data, they find consistency with $\Lambda$CDM at the ${\sim}\,1\sigma$ level.

While such model-independent approaches that reconstruct the DE density in redshift bins~\cite{Wang:2004ru,Wang:2025vfb,DESI:2025fii} offer maximal flexibility, they come at the cost of additional free parameters.
Moreover, the information content of geometric probes is largely concentrated in two effective DE degrees of freedom~\cite{Linder:2005ne,DESI:2025fii}.
This motivates the approach pursued in this work: a minimal two-parameter polynomial expansion of the normalized DE density $f_{\rm DE}(z)\equiv \rho_{\rm DE}(z)/\rho_{\rm DE,0}$ that, like CPL, requires only two free parameters.
Its smooth analytical form also lends itself naturally to the inclusion of DE perturbations, unlike the discontinuous bin structure of model-independent reconstructions.
A related approach was pursued in Ref.~\cite{Sen:2007gk} through a polynomial parameterization of the DE pressure, recently constrained using current data in~\cite{Cheng:2025lod}.
As we discuss in Sec.~\ref{sec:polynomial}, the density parameterization introduced here spans an equivalent functional space but differs crucially in the choice of parameters.

Specifically, our polynomial density expansion is parameterized in terms of the DE density fraction $f_p \equiv f_{\rm DE}(z_p)$ and equation of state $w_p \equiv w(z_p)$ at the pivot redshift.
Here $f_p$ directly quantifies how much the DE density differs from its present-day value, while $w_p$ captures its rate of evolution.
Together, they provide an immediate physical characterization of DE at the epoch where current observations are most sensitive.
Moreover, as we show in this work, these two parameters are not only largely decorrelated but are each predominantly constrained by a distinct subset of the data.
In particular, $w_p$ is determined by BAO and CMB measurements, while $f_p$ is fixed by the independent matter density constraint that SNe supply.

The remainder of this paper is organized as follows.
In Sec.~\ref{sec:2} we develop the polynomial DE density parameterization, compare it with CPL, and validate it against representative quintessence scenarios.
Section~\ref{sec:methodology} outlines the datasets and analysis framework employed in this work.
We present and discuss our results in Sec.~\ref{sec:results}, and conclude with a summary and outlook for future work in Sec.~\ref{sec:conclusions}.
Finally, in Appendix~\ref{app:single_fp} we present the data analysis of the single-parameter DE formulation obtained by truncating our density expansion at first order, while Appendix~\ref{app:validation} and~\ref{app:posteriors} include supplementary material covering validation tests against quintessence benchmarks and mock data, and comprehensive posterior distributions across all models and dataset combinations considered in this work.

\section{Minimal Density Level Dark Energy Parameterization}
\label{sec:2}

\noindent In this section we develop the minimal two-parameter formulation of the DE density evolution that constitutes the main subject of this work, and compare it with the canonical CPL parametrization of the DE equation of state.

We begin in Sec.~\ref{sec:geometric} by reviewing how DE enters the geometric observables measured by BAO, CMB, and SNe data, motivating a direct parameterization of the DE density rather than the equation of state.
In Sec.~\ref{sec:polynomial} we introduce a polynomial expansion of the normalized DE density, $f_{\rm DE}$, truncated at second order, and in Sec.~\ref{sec:pivot} we reparameterize it in terms of the DE density fraction and equation of state at a pivot redshift, yielding two largely decorrelated parameters with a direct connection to the state of DE at a definite epoch.
Finally, in Sec.~\ref{sec:validation} we validate this formulation against representative quintessence scenarios, demonstrating that it recovers the expansion history with accuracy comparable to CPL while avoiding the degeneracies and redundancies inherent to the $(w_0,\,w_a)$ basis.

\subsection{Geometric Signatures of Evolving Dark Energy}\label{sec:geometric}

\noindent The primary cosmological probes of DE constrain the expansion history of the universe through measurements of distances at various redshifts.
In a spatially flat universe the evolution of the expansion rate is governed by the Friedmann equation,
\begin{align}\label{eq:Hz}
    \frac{H(z)}{H_0}=&\Big[\Omega_{bc}\,(1+z)^3+\Omega_\gamma\,(1+z)^4+ \Omega_\nu \frac{\rho_\nu(z)}{\rho_{\nu,0}} \nonumber\\
    &+\Omega_{\rm DE}\,f_{\rm DE}(z)\Big]^{1/2},
\end{align}
where $H_0$ is the Hubble parameter today, $\Omega_{bc}= \Omega_b+\Omega_c$, and $\Omega_b,\,\Omega_c,\,\Omega_\nu,\,\Omega_\gamma$, and $\Omega_{\rm DE}$ refer to the present-time energy density parameters in baryons, cold dark matter, neutrinos, radiation, and DE, respectively.
The quantity $f_{\rm DE}(z)\equiv \rho_{\rm DE}(z)/\rho_{\rm DE,0}$ is the DE density normalized to its present-day value, with $\Omega_{\rm DE}=1-\Omega_{bc}-\Omega_\gamma-\Omega_\nu$ fixed by the flatness condition.

From $H(z)$, one can construct the cosmological distances measured by each probe.
In particular, the combination of CMB and BAO data yields absolute measurements of the comoving distance,
\begin{equation}
    D_M(z)=\int_0^z\frac{{\rm d}z^\prime}{H(z^\prime)}\,,
\end{equation}
and the Hubble distance,
\begin{equation}
    D_H(z)=1/H(z)\,.
\end{equation}
BAO surveys alone measure these quantities only relative to the sound horizon at the drag epoch, $r_d$, extracting the ratios $D_M(z)/r_d$ and $D_H(z)/r_d$ from the transverse and line-of-sight clustering of spectroscopic tracers, respectively.
CMB anisotropy data provide tight constraints on the physical baryon and cold dark matter densities, which in turn determine $r_d$ to ${\sim}0.2\%$ precision~\cite{Planck:2018vyg}, allowing these ratios to be converted into absolute distance measurements.
The CMB further measures the angular size of the sound horizon at recombination $\theta_\star = r_\star/D_M(z_\star)$, where $r_\star$ is the comoving sound horizon at recombination ($z_\star \approx 1100$), with a fractional precision of order ${\sim}10^{-4}$~\cite{Planck:2018vyg}.
This effectively provides an anchor at $z_\star$ that, together with the BAO distances at lower redshifts ($z\sim 0.3\text{--}2.3$), allows a precise reconstruction of the expansion history over a wide redshift baseline.

SNe offer a complementary low-redshift probe to these measurements by constraining the luminosity distance
\begin{equation}
    d_L(z)= (1+z)\,D_M(z),
\end{equation}
 up to an overall calibration set by the fiducial magnitude $M_B$ of SNe, assuming that they are perfectly standardized.
Even when marginalizing over this calibration, the shape of the distance-redshift relation constrains the matter fraction $\Omega_{\rm m}$ independently of $r_d$ and early-universe physics, breaking the residual degeneracies that persist in BAO and CMB data alone.\footnote{See Ref.~\cite{Loverde:2024nfi} for a detailed discussion of these geometric degeneracies, particularly in the context of neutrino mass measurements.}

As seen in Eq.~\eqref{eq:Hz}, the DE sector enters the expansion rate, and hence every geometric observable, entirely through $f_{\rm DE}$, which can be expressed in terms of its equation of state $w(z)$ via the continuity equation as
 \begin{equation}
     f_{\rm DE}=\exp\left\{ 3\,\int_0^z[1+w(z^\prime)]\frac{{\rm d}z^\prime}{1+z^\prime}\right\}\,. \label{eq:fDE}
 \end{equation}
While $f_{\rm DE}$ is the quantity that directly contributes to the expansion history and the distances derived from it, $w(z)$ enters only through this nonlinear integral mapping and is therefore harder to extract from data~\cite{Wang:2004ru, Wang:2025vfb}.
In particular, this mapping can obscure features present in $w(z)$~\cite{Maor:2000jy, Barger:2000jg} and amplify parameter uncertainties when propagated to $f_{\rm DE}$~\cite{Wang:2004ru,Wang:2025vfb}.

A canonical example is the CPL parameterization~\cite{Chevallier:2000qy,Linder:2002et}, Eq.~\eqref{eq:cpl}, for which Eq.~\eqref{eq:fDE} yields
\begin{equation}
    f_{\rm DE}(z)=(1+z)^{3(1+w_0+w_a)}\exp[-3w_a\,z/(1+z)].
\end{equation}
The normalized DE density is an exponential function of $(w_0,w_a)$, introducing a distortion in how constraints on the equation of state translate into knowledge of the DE density evolution.
On the one hand, for parameter combinations that favor a growing $f_{\rm DE}$ toward the past, uncertainties are amplified nonlinearly, growing at high redshift roughly as ${\sim}3\ln(1+z)\,\sigma(w_0+w_a)$.
On the other hand, when the preferred evolution corresponds to a decreasing $f_{\rm DE}$ in the past, the exponential mapping drives the density toward zero, artificially compressing the uncertainty bands.

More fundamentally, the nonlinear integral mapping from $(w_0,w_a)$ to $f_{\rm DE}$ is not one-to-one over the redshift range probed by observations, with extended regions of the $(w_0,w_a)$ plane oriented roughly along $\Delta w_a/\Delta w_0\approx -5$ producing nearly indistinguishable expansion histories~\cite{Shlivko:2024llw}. This approximate degeneracy is intrinsic to the $(w_0,\,w_a)$ parameterization and hence predicts the orientation of the corresponding constraint contours from observational analyses~\cite{Shlivko:2024llw}.
It can also carry the posteriors toward extreme values of $w_0$ and $w_a$ that are sensitive to the choice of priors (and their ranges) and do not necessarily correspond to physically motivated DE models (see e.g.~\cite{Steinhardt:2025znn,Bayat:2025xfr, Toomey:2025xyo}).
Hence, here we pursue a polynomial expansion of the DE density that, like CPL, requires only two free parameters, while its smooth functional form also lends itself naturally to the inclusion of DE perturbations in future work.

\subsection{Polynomial Density Evolution}\label{sec:polynomial}

\noindent The simplest, model-agnostic choice to parameterize the evolution of the DE density is a Taylor expansion around the present epoch, $a = 1$, in direct analogy with the CPL parameterization of the equation of state.
Truncating at second order gives
\begin{equation}
    f_{\rm DE}(a)=1+f_a(1-a)+f_b(1-a)^2, \label{eq:fDE_fafb}
\end{equation}
where the boundary condition $f_{\rm DE}(a=1)=1$ is satisfied by construction, and the cosmological constant is recovered for $f_a=f_b=0$.
The coefficient $f_a\equiv {\rm d}f_{\rm DE}/{\rm d}a\big|_{a=1}$ controls the linear rate of change of the DE density toward the past, with $f_a > 0$ corresponding to DE that was denser at earlier times, while $f_b\equiv {\rm d}^2f_{\rm DE}/{\rm d}a^2\big|_{a=1}$ governs the curvature of this evolution.

The corresponding equation of state follows from the continuity equation,
\begin{equation}\label{eq:wDE_fafb}
w(a) = -1 + \frac{a\bigl[f_a + 2f_b(1-a)\bigr]}{3\bigl[1 + f_a(1-a) + f_b(1-a)^2\bigr]}\,.
\end{equation}
The second term in Eq.~\eqref{eq:wDE_fafb} is proportional to $a$, so that in the early universe, as $a\rightarrow 0$, DE effectively behaves as a cosmological constant, i.e.\ $w\approx -1$, regardless of the values of $f_a$ and $f_b$.
This is a structural property of any density parameterization that is polynomial in $a$, and is consistent with a wide class of physically motivated scenarios, including thawing quintessence and K-essence models, in which the scalar field is frozen by Hubble friction at early times.
In such scenarios, the DE density deviates smoothly from its frozen value as the field begins to roll, a departure naturally captured by low-order polynomial corrections such as those in Eq.~\eqref{eq:fDE_fafb}.
A concrete realization was provided in Ref.~\cite{Sen:2007gk} in the context of a multi-field K-essence model, where each field contributes a term to the DE pressure.
This results in a polynomial expansion in powers of $(1-a)$ that, upon integration of the continuity equation, generates a polynomial density evolution of equal order.

The density parameterization presented here spans an equivalent functional space to the pressure parameterization of Ref.~\cite{Sen:2007gk}, recently constrained against the latest BAO, CMB, and SNe data~\cite{Cheng:2025lod}.
The key distinction lies in the choice of parameters.
The pressure approach uses Taylor coefficients whose values do not directly relate to physical properties of DE at any definite epoch, making the resulting constraints difficult to interpret in terms of underlying models.
As we describe in the next subsection, our density formulation can be reparameterized in terms of the DE density fraction and equation of state evaluated at a pivot scale factor, yielding two largely decorrelated parameters with immediate physical meaning that can be straightforwardly connected to predictions of quintessence and other dynamical DE models.

Before presenting this reparameterization, we note one additional property of Eq.~\eqref{eq:wDE_fafb}.
The derived equation of state diverges wherever $f_{\rm DE}$ passes through zero, which may occur for parameter values that drive the DE density negative at high redshift.
This is not a physical pathology but rather a coordinate singularity, with both the DE density and pressure remaining finite at these points.
CPL effectively exhibits the converse issue: while $w$ always varies smoothly, the derived DE density can grow or vanish exponentially for certain parameter values.
In practice, as we show in Sec.~\ref{sec:results}, when confronted with data, the constrained posteriors admit $f_{\rm DE} < 0$ only for parameter combinations in the tails of the distributions and at very early times, well beyond the epoch where DE becomes dynamically relevant and the data have meaningful constraining power.
This is consistent with the findings of Ref.~\cite{Cheng:2025lod} for the equivalent pressure parameterization.

\subsection{Pivot Parameterization}\label{sec:pivot}

\noindent As noted in the previous subsection, the Taylor coefficients $(f_a, f_b)$ that specify the density evolution in Eq.~\eqref{eq:fDE_fafb} do not directly correspond to physical properties of DE at any definite epoch, limiting the interpretability of the resulting constraints.

To address this, we reparameterize the DE density evolution in terms of the DE density fraction and equation of state at a pivot scale factor $a_p$, namely:
\begin{equation}
    f_p\equiv f_{\rm DE}(a_p), \qquad w_p\equiv w(a_p),
\end{equation}
In terms of these two quantities, the expression for the DE density ratio in Eq.~\eqref{eq:fDE_fafb} becomes:
\begin{align}
    f_{\rm DE}(z)&=f_p-3\,(1+w_p)\,f_p\left(\frac{z_p-z}{1+z}\right) \nonumber\\
    &+\frac{1-f_p+3\,(1+w_p)\,f_p\,z_p}{z_p^2}\left(\frac{z_p-z}{1+z}\right)^2,\label{eq:fDE_pv}
\end{align}
where we have rewritten the expression in terms of redshift for convenience, with $z_p = 1/a_p - 1$ denoting the corresponding pivot redshift, and the cosmological constant is recovered for $(w_p,\,f_p)=(-1,\,1)$.

The pivot redshift $z_p$ is chosen as the epoch at which a given dataset combination best constrains the DE properties, minimizing the correlation between $f_p$ and $w_p$.
This is the same concept employed in the CPL framework, where reparameterizing from $(w_0,\,w_a)$ to $(w_p, w_a)$ at a suitably chosen $z_p$ decorrelates the two parameters and minimizes the uncertainty on $w_p$~\cite{Huterer:2000mj, Albrecht:2006um, Cortes:2024lgw}, though $w_a$ itself remains a weakly constrained Taylor coefficient, offering limited additional insight into the DE dynamics at any specific epoch.
In fact, pivoting cannot resolve a structural limitation of the CPL parameterization, namely the weak sensitivity of $f_{\rm DE}$ to $w_a$ over the redshift range probed by current observations, quantified by the kernel $\partial \ln f_{\rm DE}/\partial w_a= 3\,\left[\ln(1+z)-z/(1+z)\right]$.
This insensitivity is compounded at the level of the expansion rate---the quantity that geometric probes most directly constrain---since at the higher redshifts where $\partial\ln f_{\rm DE}/\partial w_a$ is largest, DE is already subdominant to matter, so that extended ranges of $w_a$ leave the expansion history largely unaffected.

Instead, in our formulation, both parameters are individually transparent: $f_p$ directly quantifies the DE density at the pivot epoch relative to today, while $w_p$ captures its rate of evolution at that same epoch.
Together, they provide an immediate characterization of the DE density evolution that determines the cosmological distances measured by BAO, CMB, and SNe data. To summarize, the pivot achieves decorrelation in both frameworks; the distinguishing advantage of the density-level formulation is that it replaces the weakly constrained $w_a$ with $f_p$, which enters $H(z)$ directly and is therefore individually well constrained by the data.

Finally, while the focus of this work is on the two-parameter $(w_p,\,f_p)$ formulation, truncating the DE density expansion in Eq.~\eqref{eq:fDE_fafb} at first order yields an even simpler, single-parameter model, equivalently described by $f_a$ or $f_p$ alone, through the relation $f_a=(f_p-1)/(1-a_p)$.
While this minimal model captures the leading-order density evolution and performs well for single-field quintessence-like scenarios, it suffers from the same fundamental limitation as the constant equation-of-state parameterization: a single degree of freedom can only capture an effective average over the true DE evolution, and is generically driven toward the cosmological constant value, i.e. $w\approx -1$ or equivalently $f_p\approx 1$, by the CMB distance constraint~\cite{Linder:2007ka}.
We leave the data analysis of this single-parameter formulation to Appendix~\ref{app:single_fp}.

\subsection{Benchmark Quintessence Comparison with CPL}\label{sec:validation}

\begin{figure*}
    \centering
    \includegraphics[width=0.9\linewidth]{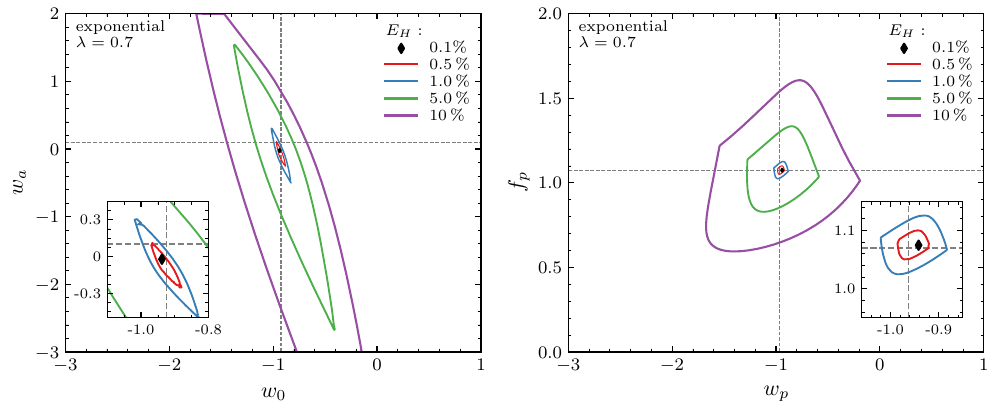}
    \caption{{\it Left panel:} Contours of the maximum relative error $E_H$, Eq.~\eqref{eq:EH}, in the $(w_p,\,f_p)$ plane for a representative exponential quintessence scenario, Eq.~\eqref{eq:exp}, with $\lambda=0.7$. Contours correspond to $E_H = 0.5\%,\,1\%,\,5\%$ and $10\%$, colored as indicated in the legend, with the black diamond marking the best-fit point corresponding to $E_H\approx 0.1\%$. Dashed \textcolor{tabgray}{gray} lines indicate the actual values of $w$ and $f_{\rm DE}$ at the pivot redshift $z_p$ for the underlying quintessence model. The inset on the lower left of the panel shows a zoomed-in view of the region enclosed by the 1\% error contour. {\it Right panel:} Same as the {\it Left panel} but for the $(w_0,\,w_a)$ parameterization, with dashed \textcolor{tabgray}{gray} lines now indicating the actual values of $w_0$ and $w_a$, defined as the zeroth- and first-order Taylor coefficients of $w(a)$ at $a=1$. The compact, non-degenerate contour structure in the $(w_p,\,f_p)$ plane, together with the near-coincidence of the model's physical properties at the pivot with the best fit, illustrates the fidelity and stability of the density-level parameterization relative to CPL.}
    \label{fig:1}
\end{figure*}

\noindent Before confronting observational data, we assess how faithfully the polynomial DE density parameterization captures the expansion history of concrete DE models, and how it compares with CPL.
As benchmarks, we consider thawing quintessence models driven by a single scalar field minimally coupled to gravity, which represent one of the simplest and best-studied realizations of dynamical DE.
In these scenarios, the field is frozen by Hubble friction at early times, behaving as an effective cosmological constant, and begins to roll only at late times as $H$ decreases, driving the equation of state above $w=-1$ and causing the DE density to dilute with the expansion.

In particular, we consider two representative classes of thawing
quintessence: exponential and hilltop potentials, which are motivated by supergravity, superstring theories~\cite{Wetterich:1994bg,Binetruy:1998rz,Bedroya:2019snp}, as well as axion-like models~\cite{Dutta:2008qn,Frieman:1995pm}, and are respectively defined as
\begin{align}
    V_{\rm exp}(\phi)&=V_0\,e^{\lambda\phi/M_{\rm pl}}, \label{eq:exp}\\
    V_{\rm hill}(\phi)&=V_0\left(1-\frac{k^2\phi^2}{2}\right).\label{eq:hill}
\end{align}
Here $M_{\rm pl}=1/\sqrt{8\pi G}\approx 2.4\times 10^{18}\,{\rm GeV}$ is the reduced Planck mass, $V_0$ is the scale of the potential, fixed by the observed DE density today, with $\lambda$ and $k$ respectively controlling the slope of the exponential and the curvature at the hilltop maximum.

To quantify the accuracy of each parameterization in reproducing the expansion history of a given quintessence scenario, we follow~\cite{Shlivko:2024llw, Toomey:2025xyo} and minimize the maximum relative error in the Hubble rate, namely\footnote{The choice to minimize $E_H$ is not unique; one could for instance adopt a root-mean-square criterion instead. We have verified that this alternative yields negligible differences in the mapped distributions and does not affect our conclusions (see also~\cite{Shlivko:2025fgv,Toomey:2025xyo}).}:
\begin{equation}
    E_H=\underset{z<4}{\rm max}\left|\frac{H_{\rm fit}(z) - H_\phi(z)}{H_\phi(z)}\right|,\label{eq:EH}
\end{equation}
where $H_\phi(z)$ is the Hubble evolution of the underlying quintessence model, computed numerically from the scalar field dynamics using \texttt{class\_ede}\footnote{\hyperlink{https://github.com/mwt5345/class_ede}{\texttt{
https://github.com/mwt5345/class\_ede}}}; $H_{\rm fit}(z)$ is the Hubble rate obtained from either the CPL or the $(w_p,\,f_p)$ parameterization; and $z < 4$ roughly encompasses the redshift range covered by current BAO and supernova measurements.\footnote{DESI BAO measurements span seven redshift bins from $z=0.3$ to $z=2.33$, with the last bin extending out to $z\approx 3.5$~\cite{DESI:2024uvr,DESI:2024aqx}. Restricting this redshift range or modifying the binning to more closely match these measurements has negligible impact on the results of this validation.}

Figure~\ref{fig:1} illustrates the contours of $E_H$ in the $(w_0,\,w_a)$ and $(w_p,\,f_p)$ planes for an exponential potential with $\lambda = 0.7$, a representative value within the theoretically motivated range~\cite{Toomey:2025xyo}.
In both parameterizations, the best-fit point achieves $E_H\approx 0.1\%$, confirming that the polynomial DE density formulation reproduces the quintessence expansion history with accuracy comparable to CPL.
Despite this, the regions of parameter space over which this accuracy is maintained differ significantly between the two bases.
In the $(w_0,\,w_a)$ plane, the $E_H$ contours are elongated along the $\Delta w_a/\Delta w_0 \approx -5$ direction discussed in the previous section, with pairs producing a maximum relative error in $H(z)$ of ${\sim}\,5\%$ already spanning most of the $w_a$ range typically adopted in observational analyses, directly reflecting the weak sensitivity of the expansion rate to $w_a$ inherent to the CPL mapping.
Moreover, the actual zeroth- and first-order Taylor coefficients of the equation of state at $a=1$ for the underlying quintessence model (dashed gray lines in the figure) fall outside the 1\% error contour, specifically giving $E_H \simeq 1.4\%$ in this example.
This is expected, as $H(z)$ depends on $w(z)$ only through an integral, Eq.~\eqref{eq:fDE}, so that the best-fit CPL parameters generically do not coincide with the local values of the equation of state and its derivative at $a=1$.

In contrast, the $E_H$ contours in the $(w_p,\,f_p)$ plane are compact and approximately orthogonal, with no extended degeneracy direction.
The actual values of the fractional DE density and equation of state at the pivot redshift for the quintessence model under consideration nearly coincide with the best fit, falling within the 0.5\% contour with $E_H \simeq 0.2\%$, confirming that the parameterization faithfully recovers the physical DE properties at the pivot epoch.
As anticipated, this reflects the fact that smooth, slowly evolving DE density evolutions of the kind produced by thawing quintessence are naturally well captured by a low-order polynomial expansion in $(1-a)$.

The same qualitative features are observed for the representative hilltop example with $k = 3/M_{\rm pl}$ and $\phi_i = 10^{-2}\,M_{\rm pl}$, shown in Fig.~\ref{fig:A1} of Appendix~\ref{app:validation}, where the best-fit accuracy improves further to $E_H \approx 0.06\%$ for both parameterizations.
As in the exponential case, the zeroth- and first-order Taylor coefficients of $w(a)$ around $a=1$ fall outside the 1\% error contour in the $(w_0,\,w_a)$ plane, while in the $(w_p,\,f_p)$ plane the actual values at the pivot coincide with the best fit even more precisely than for the exponential potential.
We have verified that these features persist across the theoretically motivated parameter ranges for both potential classes, as identified in Ref.~\cite{Toomey:2025xyo}.

Taken together, these results demonstrate that the $(w_p,\,f_p)$ parameterization can recover the DE density and equation of state at the pivot epoch, free from the degeneracies that affect the CPL basis.
To further validate our polynomial DE density formulation, we also present in Appendix~\ref{app:validation} an analysis of mock data mimicking current DESI, CMB, and SNe observations, generated from both $\Lambda$CDM and exponential quintessence cosmologies, showing an unbiased parameter recovery and consistent reconstruction of $f_{\rm DE}(z)$ and $w(z)$.

\section{Datasets and Methodology}\label{sec:methodology}

\noindent Having established the reliability of the polynomial density parameterization through the quintessence benchmarks of the previous section, we now proceed to confront it with current cosmological observations.

To this end, we perform Markov Chain Monte Carlo (MCMC) analyses using the publicly available sampler \texttt{MontePython}~\cite{Audren:2012wb, Brinckmann:2018cvx}, interfaced with a modified version of the Boltzmann solver \texttt{CLASS}~\cite{Blas:2011rf} that implements the polynomial fractional DE density of Eq.~\eqref{eq:fDE_fafb}.
All analyses are carried out at the background level only, which allows us to consistently explore regions of parameter space where $f_{\rm DE}$ becomes negative at early times, without the complications that singularities in the derived equation of state $w$ at such crossings would introduce in perturbation calculations.
We defer the inclusion of DE perturbations to future work.

\begin{figure*}
    \centering
    \includegraphics[width=0.95\linewidth]{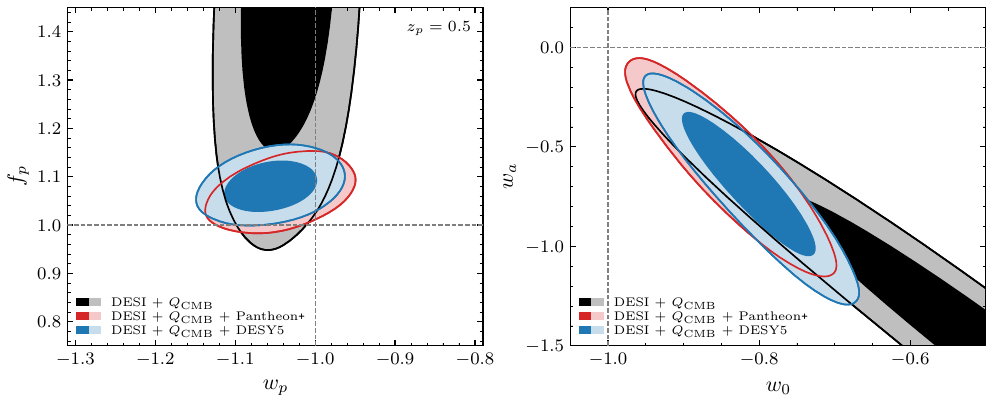}
    \caption{ {\it Left panel:} Two-dimensional posteriors in $w_p$ and $f_p$ at the 68\% and 95\% confidence levels from the DESI + $Q_{\rm CMB}$ (black), DESI + $Q_{\rm CMB}$ + Pantheon\texttt{+} (\textcolor{tabred}{red}), and DESI + $Q_{\rm CMB}$ + DESY5 (\textcolor{tabblue}{blue}) dataset combinations. The pivot redshift is set to $z_p=0.5$, which sufficiently decorrelates the two parameters across all dataset combinations shown. The \textcolor{tabgray}{gray} dashed lines mark the $\Lambda$CDM limit given by $(w_p,\,f_p) = (-1,\,1)$. {\it Right panel:} Same as the {\it Left panel} but for the $(w_0,\,w_a)$ parameterization, with dashed \textcolor{tabgray}{gray} lines indicating again the $\Lambda$CDM limit, which here corresponds to $(w_0,\,w_a) =(-1,\,0)$. The inclusion of SNe dramatically tightens the constraints in both parameterizations, increasing the preference for evolving DE while simultaneously shifting all posteriors toward their $\Lambda$CDM values, with the $(w_p,\,f_p)$ formulation delivering percent-level measurements of both DE parameters individually.}
    \label{fig:2}
\end{figure*}

For the baseline $\Lambda$CDM model, we vary the physical baryon density $\omega_b$, the physical cold dark matter density $\omega_{\rm cdm}$, and the reduced Hubble constant $h\equiv H_0/(100\,{\rm km}\,{\rm s}^{-1}\,{\rm Mpc}^{-1})$, adopting the same flat priors as the DESI collaboration (see Table I in~\cite{DESI:2025fii}).
In the CPL extension, we additionally vary $w_0\in[-3,\,1]$ and $w_a \in [-3,\,2]$ with the condition $w_0+w_a<0$ to enforce a period of high-redshift matter domination, again following the DESI baseline.
For our density parameterization, we set broad flat priors, varying $w_p \in [-3,\,1]$ and $f_p \in [-1,\,3]$, and fix the pivot redshift to $z_p=0.5$ for all analyses, as done in~\cite{Efstathiou:2025tie}.
In principle, $z_p$ could be optimized individually for each dataset combination to fully decorrelate $w_p$ and $f_p$, analogous to the standard procedure in the CPL framework~\cite{Huterer:2000mj, Albrecht:2006um} (see also~\cite{Herold:2025hkb}), but as shown in Fig.~\ref{fig:2} this choice achieves a reasonable decorrelation across all data combinations considered here.

Since our analyses do not include perturbation-level constraints, we fix the remaining $\Lambda$CDM parameters to the same {\it Planck}-inspired values adopted by the DESI collaboration for their background-only analyses~\cite{DESI:2025zgx, DESI:2025fii}, namely $\ln(10^{10}A_s) = 3.036$, $n_s = 0.9649$, and $\tau_{\rm reio} = 0.0544$.
We additionally assume three degenerate massive neutrino species with $\sum m_\nu = 0.06\,{\rm eV}$ and $N_{\rm eff} = 3.044$.

We use \texttt{GetDist}~\cite{2019arXiv191013970L} to analyze all MCMC chains, which are run until the Gelman--Rubin convergence criterion~\cite{Gelman:1992zz} $\hat{R}-1<0.01$ is satisfied.

\smallskip

We consider the following datasets in our analyses:\footnote{The public likelihoods for the BAO and SNe data considered in this work were originally developed for the \texttt{Cobaya} sampler~\cite{2019ascl.soft10019T, Torrado:2020dgo}. Here we use the \texttt{MontePython}-compatible implementations of the DESI DR2 BAO, as well as the Pantheon\texttt{+} and DESY5 SNe likelihoods, respectively from Ref.~\cite{Herold:2024nvk} and~\cite{Herold:2025hkb}.}

\begin{itemize}
    \item DESI DR2 BAO: We use the BAO data from the second DESI data release~\cite{DESI:2025zgx} (DR2), which include isotropic and anisotropic distance measurements from galaxy, quasar, and Lyman-$\alpha$ tracers spanning the redshift range $0.1<z<4.2$ across seven principal redshift bins.
    We refer to this dataset as `DESI'.

    \item Compressed CMB: We adopt the Gaussian correlated prior on $\omega_b$, $\omega_{bc} \equiv \omega_b + \omega_c$, and $\theta_\star$ as defined in Ref.~\cite{DESI:2025zgx} (specifically Eqs.~(A1) and (A2) in Appendix A).
    This compressed likelihood captures the geometrically relevant information from the CMB, i.e., the calibration of the sound horizon $r_d$ through the physical baryon and cold dark matter densities and the distance to last scattering through $\theta_\star$, while marginalizing over late-time effects such as the integrated Sachs--Wolfe effect and CMB lensing, making it well suited for probing late-time physics independently of assumptions about DE evolution~\cite{Lemos:2023xhs}.
    We refer to this dataset as `$Q_{\rm CMB}$', and always use it in conjunction with DESI.

    \item Type Ia Supernovae: We consider two supernova samples.
    The first is Pantheon\texttt{+}~\cite{Scolnic:2021amr,Brout:2022vxf}, comprising 1701 light curves from 1550 distinct SNe over the redshift range $0.001 < z < 2.26$.
    The second is the full five-year dataset from the Dark Energy Survey (DESY5)~\cite{DES:2024jxu, DES:2024hip, DES:2024upw}, containing 1635 SNe in the range $0.1 < z < 1.13$, in the updated Dovekie calibration~\cite{Popovic:2025glk, DES:2025sig} that corrects systematic calibration errors identified in the original release.
    We use these supernova samples only in conjunction with DESI + $Q_{\rm CMB}$, and respectively refer to them as `Pantheon\texttt{+}' and `DESY5'.
    For brevity, we do not include the Union3 sample~\cite{Rubin:2023jdq}, which has substantial overlap with Pantheon\texttt{+}, albeit with a markedly different analysis framework.
    All three SNe compilations have been shown to yield consistent constraints in the context of evolving DE~\cite{Hoyt:2026fve}, and we therefore expect the conclusions drawn here to hold broadly for Union3 as well.
\end{itemize}

In summary, we analyze three data combinations: (i) DESI + $Q_{\rm CMB}$, (ii) DESI + $Q_{\rm CMB}$ + Pantheon\texttt{+}, and (iii) DESI + $Q_{\rm CMB}$ + DESY5.

To quantify the statistical preference for evolving DE, we compute $\Delta \chi^2_{\rm MAP}$, defined as twice the difference in the maximum log-likelihood between $\Lambda$CDM and each two-parameter DE extension.
Since $\Lambda$CDM is nested within both the CPL and $(w_p,\,f_p)$ parameterizations, Wilks' theorem~\cite{Wilks:1938dza} implies that, under the null hypothesis and assuming Gaussian errors, $\Delta\chi^2_{\rm MAP}$ follows a $\chi^2$ distribution with two degrees of freedom.
To express this preference in more intuitive units, we quote the corresponding frequentist significance $N\sigma$ obtained by matching the $p$-value of the $\chi^2$ test to the two-tailed Gaussian tail probability,
\begin{equation}
N = \sqrt{2}\;\mathrm{erfc}^{-1}\!\left(e^{-\Delta\chi^2_{\rm MAP}/2}\right),
\end{equation}
where $\mathrm{erfc}^{-1}$ denotes the inverse of the complementary error function, $\mathrm{erfc}(x) = 2/\sqrt{\pi}\,\int_x^{\infty} \exp({-t^2})\,dt$.

\section{Results \& Discussion}\label{sec:results}

\noindent We now present the results of our analysis, constraining the $(w_p,\,f_p)$ DE density parameterization with the three data combinations described in the previous section and comparing it directly with CPL as a benchmark.
We begin by examining the parameter constraints and the complementary roles played by each dataset, then discuss the reconstructed DE density and equation-of-state evolution, and conclude by examining the statistical preference for evolving DE over $\Lambda$CDM and its robustness to prior choices and statistical framework.

\subsection{Constraints and Dataset Complementarity}

\noindent Our primary results are shown in Figure~\ref{fig:2}, which displays the posterior distributions in the $(w_p,\,f_p)$ plane from the three data combinations under consideration, with the corresponding CPL results shown in the right panel for comparison.
Table~\ref{tab:1} accompanies these posteriors, summarizing the 68\% confidence level (C.L.) constraints on the key cosmological parameters varied in each analysis. The full posterior distributions, along with supplementary reconstructions and model comparisons, are collected in Appendix~\ref{app:posteriors}.

With DESI + $Q_{\rm CMB}$ alone, the $(w_p,\,f_p)$ parameterization yields:
\begin{equation}
\left.
\begin{aligned}
w_p &= -1.03\pm 0.04   \\
f_p &= 1.66^{+0.41}_{-0.23}
\end{aligned}
\ \right\} \quad \substack{\text{\normalsize DESI + $Q_{\rm CMB}$ }.}
\end{equation}
The 68\% C.L. on both $\Omega_{\rm m}$ and $H_0$ widen by roughly an order of magnitude compared to the $\Lambda$CDM analysis, with their mean values also shifting ${\sim}\,2\sigma$ above and below their $\Lambda$CDM values, respectively.
This broadening reflects a fundamental limitation of BAO and CMB data in the presence of evolving DE.
Both probes constrain distances normalized to a fixed physical scale---the sound horizon $r_d$ for BAO, and the angular size of the sound horizon at decoupling $\theta_\star$ for the CMB---and are therefore primarily sensitive to how $H(z)$ varies across redshift, but less so to its overall normalization, leaving a significant degeneracy between the total DE density and the parameters governing the expansion rate today.

The CPL analysis of the same data exhibits qualitatively similar behavior, returning
\begin{equation}
\left.
\begin{aligned}
w_0 &= -0.42^{+0.21}_{-0.24}   \\
w_a &= -1.73^{+0.68}_{-0.62} 
\end{aligned}
\ \right\} \quad \substack{\text{\normalsize DESI + $Q_{\rm CMB}$ },}
\end{equation}
consistent with the analogous constraints reported by the DESI collaboration~\cite{DESI:2025zgx}.\footnote{We note a minor $\lesssim 0.1\sigma$ difference in our $w_0w_a$CDM constraints from DESI + $Q_{\rm CMB}$ relative to those reported by the DESI collaboration~\cite{DESI:2025zgx}, likely attributable to the different choices of sampler (\texttt{MontePython} vs.\ \texttt{Cobaya}~\cite{2019ascl.soft10019T,Torrado:2020dgo}), Boltzmann solver (\texttt{CLASS} vs.\ \texttt{CAMB}~\cite{Lewis:1999bs}), and neutrino mass splitting (three degenerate species vs.\ one massive and two massless).}
The structure of these degeneracies differs, however, between the two parameterizations.
In CPL, both $w_0$ and $w_a$ contribute to the overall normalization of $H(z)$ through the exponential mapping of Eq.~\eqref{eq:fDE}, and are accordingly both correlated with $H_0$ and $\Omega_{\rm m}$, producing the well-known elongated contours along $\Delta w_a/\Delta w_0 \approx -5$ in the $(w_0,\,w_a)$ plane~\cite{Shlivko:2024llw}.
In the $(w_p,\,f_p)$ parameterization, this degeneracy is largely absorbed by $f_p$, which directly controls the DE density at the pivot redshift and is therefore tied to the overall normalization of the expansion rate, as reflected in its broad and asymmetric posterior (see also~Fig.~\ref{fig:A6}).
The equation of state $w_p$, on the other hand, governs the rate of change of the DE density at the pivot epoch, and is thus constrained by the relative variation of $H(z)$ across the BAO redshift bins anchored by the CMB distance to last scattering, which already suffices to determine it to percent-level precision.

The inclusion of SNe shifts and significantly tightens the $(w_p,\,f_p)$
constraints. With DESI + $Q_{\rm CMB}$ + DESY5, we obtain
\begin{equation}
\left.
\begin{aligned}
w_p &= -1.06\pm 0.04   \\
f_p &= 1.08\pm 0.04
\end{aligned}
\ \right\} \quad \begin{aligned}
&\text{\normalsize DESI + $Q_{\rm CMB}$ } \\
&\text{\normalsize + DESY5.}
\end{aligned}
\end{equation}
Relative to the DESI + $Q_{\rm CMB}$ only analysis, the uncertainty on $f_p$ shrinks by nearly an order of magnitude, a direct consequence of SNe measurements constraining $\Omega_{\rm m}$ through the shape of the luminosity distance--redshift relation independently of $r_d$, thereby breaking the degeneracy between the DE density and the overall normalization of the expansion rate identified above.
The uncertainties on $\Omega_{\rm m}$ and $H_0$ accordingly tighten back to values comparable to those obtained in our $\Lambda$CDM analyses, despite the two additional parameters, and notably, the mean values of all cosmological parameters also shift significantly toward their $\Lambda$CDM values (see Table~\ref{tab:1}).

Replacing DESY5 with Pantheon\texttt{+} yields consistent results, with the posteriors shifting marginally closer to $\Lambda$CDM, in particular yielding
\begin{equation}
\left.
\begin{aligned}
w_p &= -1.04\pm 0.04   \\
f_p &= 1.07\pm 0.04
\end{aligned}
\ \right\} \quad \begin{aligned}
&\text{\normalsize DESI + $Q_{\rm CMB}$ } \\
&\text{\normalsize + Pantheon\texttt{+}.}
\end{aligned}
\end{equation}

The CPL constraints undergo a qualitatively similar sharpening upon the inclusion of SNe data, with the uncertainties on $w_0$ and $w_a$ reduced by roughly a factor of four and three, respectively, while both central values also move appreciably toward their cosmological constant limits.

Beyond these broad similarities, however, the two parameterizations differ markedly in how the constraining power is distributed among their DE parameters.
In the $(w_p,\,f_p)$ parameterization, both DE parameters are individually measured to percent-level precision, delivering a direct and robust characterization of the DE density fraction and equation of state at the pivot epoch that is stable across the two SNe samples considered here.
The CPL parameters, by contrast, remain significantly correlated even after the inclusion of SNe, with the residual degeneracy spreading the available information across the two parameters rather than localizing it in each individually.
One may of course reparameterize CPL as $(w_p,\,w_a)$, which by construction decorrelates the two parameters and yields a comparably tight constraint on $w_p$ (see e.g.~\cite{Cortes:2024lgw, Efstathiou:2025tie}).
However, as discussed in Sec.~\ref{sec:2}, the relatively broad uncertainties of $w_a$ are a structural consequence of the weak sensitivity of the expansion rate to this parameter, which therefore offers only modest additional insight into the DE dynamics at any specific redshift.

The $(w_p,\,f_p)$ formulation goes beyond this simple decorrelation.
As shown above, each parameter is not only statistically independent of the other, but is predominantly constrained by a distinct subset of the data.
The pivot redshift is chosen as the epoch where the data best constrain DE, and at that epoch the rate of DE density evolution, $w_p$, is directly determined by how distances vary across the BAO redshift bins relative to the CMB anchor, while the DE density fraction, $f_p$, requires the independent $\Omega_{\rm m}$ measurement that only SNe provide.
The two parameters are thus naturally aligned with the distinct observables delivered by each dataset, which allows the combined analysis to optimally constrain both of them.

Taken at face value, the constraints obtained here point to a DE density at the pivot epoch that is a few percent higher than its present-day value, with an equation of state consistent with---but mildly below---the cosmological constant limit.

\begin{table*}
	\centering
	\begin{tabular}{l @{\hskip 2em} c @{\hskip 2em} c @{\hskip 1em} c @{\hskip 1em} c}
			\toprule

	Model/Dataset & {$\Omega_{\rm m}$} & {$H_0\,[{\rm km}\,{\rm s}^{-1}\, {\rm Mpc}^{-1}]$} & {$w_0$} or $w_p$ & {$w_a$} or $f_p$\\
            \midrule[0.065em]
    $\mathbf{\Lambda}{\rm \bf CDM}$ & & & &\\
    [4pt]
	DESI + $Q_{\rm CMB}$ & $0.300 \pm 0.004$ & $68.3 \pm 0.3$ &---& ---	\\
    [4pt]
    DESI + $Q_{\rm CMB}$ + Pantheon\texttt{+} & $0.302 \pm 0.004$ & $68.2 \pm 0.3$	&---& ---	\\[4pt]
    DESI + $Q_{\rm CMB}$ + DESY5 & $0.302 \pm 0.004$ & $68.2 \pm 0.3$	&---& ---	\\[2pt]
    \midrule[0.065em]
    $\boldsymbol{w_0w_a}{\rm \bf CDM}$ & & & &\\
    [4pt]
	DESI + $Q_{\rm CMB}$ & $0.352 \pm 0.022$ & $63.6^{+2.2}_{-1.7}$ 	& $-0.42^{+0.21}_{-0.24} $ &	$-1.73^{+0.68}_{-0.62} $\\
    [4pt]
    DESI + $Q_{\rm CMB}$ + Pantheon\texttt{+} & $0.311 \pm 0.006$ & $67.4\pm 0.6$ 	& $-0.84\pm 0.06$ &	$-0.59^{+0.21}_{-0.24} $	\\[4pt]
    DESI + $Q_{\rm CMB}$ + DESY5 & $0.313 \pm 0.006$ & $67.3\pm 0.6$ 	& $-0.81\pm 0.06$ &	$-0.69^{+0.22}_{-0.25} $	\\[2pt]
    \midrule[0.065em]
    $\boldsymbol{w_p f_p}{\rm \bf CDM}$ & & & &\\
    [4pt]
	DESI + $Q_{\rm CMB}$ & $0.387^{+0.045}_{-0.032}$ & $60.9^{+2.8}_{-3.1}$ 	& $-1.03\pm 0.04 $ &	 $1.66^{+0.41}_{-0.23} $\\
    [4pt]
    DESI + $Q_{\rm CMB}$ + Pantheon\texttt{+} & $0.311\pm 0.006$ & $67.4\pm 0.6$ 	& $-1.04\pm 0.04 $ &	 $1.07\pm 0.04 $	\\[4pt]
    DESI + $Q_{\rm CMB}$ + DESY5 & $0.317\pm 0.006$ & $67.2\pm 0.6$ 	& $-1.06\pm 0.04 $ &	 $1.08\pm 0.04$ 	\\[2pt]
	\bottomrule
	\end{tabular}
	\caption{Marginalized posterior means and 68\% C.L. constraints on the key cosmological parameters from DESI DR2 BAO (labeled as `DESI') in combination with the compressed CMB likelihood $Q_{\rm CMB}$ and additional Type Ia supernovae (SNe) datasets, namely Pantheon\texttt{+} and DESY5. Results are shown for the $\Lambda$CDM, $w_0w_a$CDM, and $w_pf_p$CDM models, reporting $\Omega_{\rm m}$, $H_0$, and the corresponding DE parameters in each case. Across both two-parameter extensions, the inclusion of SNe substantially tightens the constraints and shifts all parameters toward their $\Lambda$CDM values, with the $(w_p,\,f_p)$ parameterization delivering percent-level measurements of both DE parameters individually.}
	\label{tab:1}
\end{table*}

\subsection{Reconstructed Dark Energy Evolution}
\begin{figure}
    \centering
    \includegraphics[width=\linewidth]{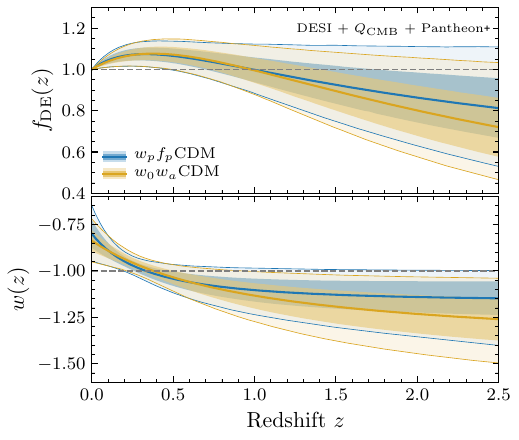}
    \caption{
     Normalized dark energy density, $f_{\rm DE}(z) \equiv
\rho_{\rm DE}(z)/\rho_{\rm DE,0}$ ({\it Top panel}), and corresponding equation of state parameter, $w(z)$ ({\it Bottom panel}), as a function of redshift from the DESI + $Q_{\rm CMB}$ + Pantheon\texttt{+} analysis. Results are shown for the $(w_p,\,f_p)$ (\textcolor{tabblue}{blue}) and $(w_0,\,w_a)$ (\textcolor{goldenrod}{gold}) parameterizations, with solid lines indicating the posterior median and increasingly lighter shading denoting the 68\% and 95\% confidence level bands. The dashed gray line marks the $\Lambda$CDM limit, namely $f_{\rm DE} = 1$ and $w=-1$. Both parameterizations yield consistent reconstructions that remain close to the cosmological constant limit, with a DE density that rises mildly above unity at intermediate redshifts and an equation of state that remains marginally phantom over the range where the data have constraining power.    }
    \label{fig:3}
\end{figure}

\noindent Figure~\ref{fig:3} shows the reconstructed DE density evolution $f_{\rm DE}(z)$ and equation of state $w(z)$ from the DESI + $Q_{\rm CMB}$ + Pantheon\texttt{+} analysis, which yields the more conservative constraints of the two SNe samples considered here and is therefore shown in the main text, with the corresponding DESY5 results presented in Fig.~\ref{fig:A2} in the Appendix.
In both panels, the $(w_p,\,f_p)$ and CPL reconstructions are overlaid, with the 68\% and 95\% C.L. bands derived from the respective posterior distributions.

The two parameterizations yield consistent reconstructions across the redshift range probed by the data, providing an important crosscheck that both formulations capture the same underlying expansion history despite their distinct functional forms.
Both favor a DE density that rises a few percent above its present-day value by $z \sim 0.5$ before decreasing at higher redshifts, with an equation of state that remains mildly phantom over the range where the data have constraining power.

As discussed in Sec.~\ref{sec:2}, a direct parameterization of the DE density propagates parameter uncertainties into $f_{\rm DE}(z)$ as polynomial corrections, whereas equation-of-state parameterizations such as CPL do so through the exponential mapping of Eq.~\eqref{eq:fDE}, which can amplify uncertainties nonlinearly at high redshift.
This advantage is present in the reconstructions shown here, with the $(w_p,\,f_p)$ bands displaying a controlled, polynomial growth of uncertainties with redshift.
However, the practical gain over CPL is more modest than one might naively expect, because the current data favor a DE density that turns over at moderate redshift rather than growing monotonically toward the past.
This particular shape of the evolution keeps the CPL exponential mapping in a regime where the amplification of uncertainties remains mild, resulting in $f_{\rm DE}(z)$ and $w(z)$ bands of comparable width in the two parameterizations.
As shown by the mock quintessence analyses of Appendix~\ref{app:validation}, this difference becomes considerably more pronounced when the underlying DE density grows monotonically toward the past and the nonlinear amplification of uncertainties inherent to the CPL mapping is no longer tempered by the shape of the preferred evolution.

A more subtle distinction between the two reconstructions concerns the behavior at high redshift, where the CPL equation of state drifts to mildly more phantom values with a correspondingly lower $f_{\rm DE}$, while the $(w_p,\,f_p)$ reconstruction remains closer to the cosmological constant limit, displaying slightly less variation in $w(z)$ over the same range.
This systematic difference, consistently at the ${\sim}\,0.5\sigma$ level, is driven by the broad $w_a$ posterior, which extends to relatively large negative values even after the inclusion of SNe.
As discussed in Sec.~\ref{sec:2}, this breadth likely reflects the intrinsic redundancy of the CPL mapping rather than a genuine data preference.
In fact, the nonlinear relation between $(w_0,\,w_a)$ and the expansion history allows extended ranges of $w_a$ to produce effectively indistinguishable $H(z)$ over the redshift range probed by observations (see e.g.\ Fig.~\ref{fig:1}), spreading the posterior and pulling the reconstructed $w(z)$ to more phantom values at early times.
This ambiguity is absent in the $(w_p,\,f_p)$ parameterization, where both parameters are individually well constrained and directly tied to the DE density at the pivot epoch.
It is therefore noteworthy that even in this formulation, where the mapping between the DE parameters and the expansion history is considerably less degenerate, a marginal hint of phantom behavior persists across all data combinations considered here.
This further suggests that the phantom preference, while modest, is rooted in the data rather than driven by any particular choice of DE parameterization.

Finally, the corresponding DESI + $Q_{\rm CMB}$ + DESY5 reconstructions, shown in Fig.~\ref{fig:A2}, display the same qualitative features discussed above, with a marginally more pronounced departure from the cosmological constant limit in both $f_{\rm DE}(z)$ and $w(z)$.

\subsection{On the Statistical Preference for Evolving Dark Energy}

\begin{table}
	\centering
	\begin{tabular}{l c @{\hskip 2em} c}
			\toprule
	Dataset &  $\Delta\chi^2_{\rm MAP}$ & Significance\\
            \midrule[0.065em]
    $\boldsymbol{w_0w_a}{\rm \bf CDM}$ & &\\
    [4pt]
	DESI + $Q_{\rm CMB}$ 	& -8.72 & $2.5\,\sigma$	\\
    [4pt]
    DESI + $Q_{\rm CMB}$ + Pantheon\texttt{+} 	& -8.67 &	$2.5\,\sigma$\\[4pt]
    DESI + $Q_{\rm CMB}$ + DESY5	& -11.3	 & $2.9\,\sigma$\\[2pt]
    \midrule[0.065em]
    $\boldsymbol{w_p f_p}{\rm \bf CDM}$ & &\\
    [4pt]
	DESI + $Q_{\rm CMB}$  	& -8.92 & $2.5\,\sigma$	\\
    [4pt]
    DESI + $Q_{\rm CMB}$ + Pantheon\texttt{+} 	& -7.14 & $2.2\,\sigma$	\\[4pt]
    DESI + $Q_{\rm CMB}$ + DESY5 	& -10.7 & $2.8\,\sigma$\\[2pt]
	\bottomrule
	\end{tabular}
	\caption{Difference in $\chi^2_{\rm MAP}$ between the best-fit $\Lambda$CDM model and the $w_0w_a$CDM and $w_pf_p$CDM extensions, for each of the dataset combinations considered in this work. The $\chi^2_{\rm MAP}$ is defined as twice the negative log-likelihood evaluated at the best-fit point. The corresponding frequentist significance levels, computed assuming two extra free parameters as described in Sec.~\ref{sec:methodology}, are listed alongside each $\Delta\chi^2_{\rm MAP}$ value. Both parameterizations yield comparable significance across all dataset combinations, with the preference for evolving DE remaining below the $3\sigma$ level throughout.}
	\label{tab:2}
\end{table}

\noindent Having established the constraints and their physical interpretation, we now turn to the statistical preference for evolving DE over $\Lambda$CDM.
Table~\ref{tab:2} summarizes the frequentist significance of this preference, quantified through $\Delta\chi^2_{\rm MAP}$ between $\Lambda$CDM and the two DE parameterizations under consideration, i.e., $(w_p,\,f_p)$ and $(w_0,\,w_a)$.
Across all data combinations, the two yield equivalent fits, with their $\Delta\chi^2_{\rm MAP}$ differing by at most ${\sim}\,1.5$.

This preference for evolving DE over $\Lambda$CDM remains below the $3\sigma$ level across all of our analyses, ranging from $2.2\sigma$ to $2.9\sigma$ depending on the data combination and parameterization.
We note that these results are obtained at the background level only, without evolving DE perturbations through the Boltzmann equations.
In the $w_0w_a$CDM context, the inclusion of perturbations together with the full CMB likelihood---encompassing temperature, polarization, and lensing data in place of the compressed $Q_{\rm CMB}$ prior---has been shown to increase the significance marginally, reaching $3.2\sigma$ for DESI + CMB + DESY5~\cite{DES:2025sig}; we expect a comparable shift for the $(w_p,\,f_p)$ parameterization but defer this extension to future work.

Even taking these numbers at face value, several considerations warrant caution in their interpretation.
As emphasized in~\cite{Efstathiou:2025tie}, the $\Delta\chi^2_{\rm MAP}$ statistic converted to an equivalent number of $\sigma$ can give a misleading impression of the actual odds against $\Lambda$CDM, as it cannot account for the relative plausibility of the model being tested nor for the role of prior assumptions.
More broadly, in the ${\sim}\,2\text{--}3\sigma$ regime where current constraints reside, the interpretation of the preference is inherently sensitive to the choice of priors on the DE parameters~\cite{Steinhardt:2025znn,Bayat:2025xfr,Toomey:2025xyo,Toomey:2025yuy, Shlivko:2025krk}.

As a concrete illustration, Ref.~\cite{Toomey:2025xyo} showed that replacing uniform priors on $(w_0,\,w_a)$ with theory-informed priors derived from representative quintessence models reduces the preference for evolving DE by up to $2\sigma$ depending on the dataset combination considered.
The underlying mechanism is that uniform priors assign substantial volume to regions of parameter space that are physically implausible yet yield effectively equivalent expansion rates over the observed redshift range, biasing the posterior away from $\Lambda$CDM.
In the $(w_p,\,f_p)$ parameterization, where the mapping between the DE parameters and the expansion history is considerably less degenerate as demonstrated throughout this work, this sensitivity to priors is expected to be mitigated.
Nevertheless, a dedicated study analogous to that of Ref.~\cite{Toomey:2025xyo} would be needed to verify this quantitatively, which we leave to future work.

An alternative and in many respects more principled approach to model comparison is provided by the Bayesian evidence, which naturally penalizes model complexity through the integration of the likelihood over the full prior volume~\cite{Trotta:2008qt}.
Because the prior volume grows with each additional parameter, the improvement in fit must be substantial enough to overcome this built-in Occam penalty, making the Bayesian evidence a more conservative measure of model preference.
In the context of $w_0w_a$CDM, recent Bayesian analyses~\cite{Ong:2025utx, Ong:2026tta} have found that this penalty is sufficient to eliminate the frequentist preference, with the Bayes factor modestly favoring $\Lambda$CDM across all dataset combinations considered in this work
(when assuming the same priors as DESI). Since the $(w_p,\,f_p)$ parameterization yields a comparable $\Delta\chi^2_{\rm MAP}$ preference for evolving DE as $w_0w_a$CDM, we expect the Bayesian evidence to similarly (mildly) favor $\Lambda$CDM, although the precise impact of the Occam penalty will depend on the choice of priors for $w_p$ and $f_p$.

This contrast with the $\Delta\chi^2_{\rm MAP}$-based significances illustrates how the assessment of the evidence can shift considerably depending on the statistical framework and assumptions employed.
While the Bayesian approach makes the role of priors explicit and transparent, the choice of such prior distributions remains non-unique, particularly for DE parameters where theoretical guidance is limited~\cite{Efstathiou:2008ed}, and so neither framework can be regarded as delivering a definitive verdict at the current level of statistical precision.

Beyond these statistical considerations, the constraints themselves exhibit features that invite further scrutiny.
As pointed out by Refs.~\cite{Cortes:2024lgw, Efstathiou:2025tie}, reparameterizing CPL in terms of $(w_p,\,w_a)$ reveals that the equation of state at the pivot is remarkably close to $-1$, with effectively all the evidence for evolving DE concentrated in the weakly constrained slope $w_a$.
That the equation of state happens to equal the cosmological constant value precisely at the epoch where the data are most sensitive has been identified as a new coincidence problem for evolving DE interpretations~\cite{Cortes:2024lgw}.
Our results highlight this observation from a complementary perspective.

The obtained equation of state at the pivot, $w_p$, remains consistent with the cosmological constant limit to a precision of around $\pm 0.04$ across all data combinations.
In addition, once SNe data are included, the DE density fraction $f_p$ at the pivot is also measured with comparable precision and found to lie only a few percent above unity, consistent with no evolution at the $2\sigma$ level.
This need not have been the case; even with $w_p \approx -1$, which within an evolving DE interpretation would signal a phantom crossing at the pivot epoch, there is no a priori reason for the DE density---which peaks at the crossing---to lie so close to its present-day value, as it would identically for a cosmological constant.
In this sense, the proximity of both $w_p$ and $f_p$ to their $\Lambda$CDM values reinforces the coincidence noted above.

More broadly, the inclusion of SNe shifts not only the DE parameters but all cosmological parameters toward their $\Lambda$CDM values (see Table~\ref{tab:1}).
The $(w_p,\,f_p)$ parameterization makes this pattern particularly transparent, with $f_p -1$ shifting by nearly an order of magnitude in its mean value, from~${\sim}\,0.66$ to~${\sim}\,0.07\text{--}0.08$, and the 95\% C.L. contours of the combined analyses no longer overlapping with the 68\% C.L. region from DESI + $Q_{\rm CMB}$ alone, which favored substantially larger values of $f_p$, as visible in Fig.~\ref{fig:2}.

Taken together, the current evidence for evolving DE remains suggestive but not yet compelling.
The best-fit values migrate toward $\Lambda$CDM precisely when the most constraining datasets are included, yet the frequentist significance mildly increases because the posteriors contract faster than the central values approach the null hypothesis.
In this weakly informative regime, neither the choice of priors on the DE parameters nor the a priori plausibility of a given model enjoys a clear consensus, inviting caution in quoting absolute numbers for the model comparison.
The $(w_p,\,f_p)$ formulation contributes to this assessment by providing a framework in which both DE parameters are individually well measured and carry direct physical meaning, making the proximity of current constraints to the cosmological constant limit and the coincidences it entails all the more evident and worthy of further investigation.

\section{Conclusions \& Outlook}\label{sec:conclusions}

\noindent In this work we have introduced a minimal two-parameter formulation of the dark energy (DE) density evolution, constructed as a second-order polynomial expansion of the normalized DE density $f_{\rm DE}(a) \equiv \rho_{\rm DE}(a)/\rho_{\rm DE,0}$ in powers of $(1-a)$, and reparameterized in terms of the DE density fraction $f_p$ and equation of state $w_p$ at a pivot redshift $z_p$, Eq.~\eqref{eq:fDE_pv}.
The pivot is chosen as the epoch where a given dataset combination best constrains the DE properties, and at that epoch $f_p$ directly quantifies how much the DE density differs from its present-day value while $w_p$ captures its rate of evolution.
Together, they provide an immediate and transparent characterization of the state of DE at the redshift where current data are most constraining. Parameterizing the DE density directly, rather than the equation of state, avoids the uncertainty amplification and degeneracies inherent to the nonlinear integral mapping from $w(z)$ to $f_{\rm DE}$, Eq.~\eqref{eq:fDE}, through which the equation of state enters the expansion rate and every geometric observable. While model-independent reconstructions of $f_{\rm DE}$ in redshift bins offer maximal flexibility in this regard, geometric probes are largely sensitive to only two effective DE degrees of freedom~\cite{Linder:2005ne, DESI:2025fii}, motivating the minimal polynomial description pursued here.

We validated the $(w_p,\,f_p)$ formulation against representative thawing quintessence scenarios, showing that it reproduces the corresponding expansion history with accuracy comparable to CPL while circumventing the intrinsic redundancies of the $(w_0,\,w_a)$ basis, which pivoting to $(w_p,\,w_a)$ does not resolve.
As shown in Fig.~\ref{fig:1}, in the $(w_p,\,f_p)$ plane the error contours are compact and non-degenerate, and the best-fit parameters recover the actual physical properties of DE at the pivot epoch, in contrast with CPL, where the weak sensitivity of the expansion rate to $w_a$ leaves this parameter poorly constrained and limits its ability to characterize the DE dynamics at any specific redshift.

Confronting the $(w_p,\,f_p)$ parameterization with DESI DR2 BAO data, a compressed CMB likelihood ($Q_{\rm CMB}$), and Type Ia supernovae (SNe) in a background-level analysis, we have shown that this formulation delivers percent-level measurements of both DE parameters individually, with $w_p = -1.04 \pm 0.04$ and $f_p = 1.07 \pm 0.04$ for DESI + $Q_{\rm CMB}$ + Pantheon\texttt{+} at $z_p=0.5$, and consistent results with the DESY5 SNe sample.
These constraints yield fits of similar quality to CPL across all dataset combinations, as reflected both in the $\Delta\chi^2_{\rm MAP}$ values of Table~\ref{tab:2} and in the reconstructed evolution of the fractional DE density, $f_{\rm DE}(z)$, and its equation of state, $w(z)$, displayed in Figs.~\ref{fig:3} and~\ref{fig:A2}.

The key distinction lies not in the constraining power but in how that information is organized, as clearly illustrated by the posterior distributions of Fig.~\ref{fig:2}. Even in the pivoted CPL basis $(w_p,\,w_a)$, the weak sensitivity of the expansion rate to $w_a$ leaves this parameter poorly constrained, so that the constraining power of the data is concentrated almost entirely in a single well-measured parameter, $w_p$.
The $(w_p,\,f_p)$ formulation distributes this power across two individually well-measured quantities, each aligned with a distinct subset of the data.
Specifically, $w_p$ is determined to percent-level precision by DESI + $Q_{\rm CMB}$ alone, through the relative variation of distances across the BAO redshift bins anchored by the CMB, while $f_p$---which directly controls the DE density at the pivot and is therefore tied to the overall normalization of the expansion rate---is anchored by the independent $\Omega_{\rm m}$ measurement that only SNe deliver.

Taken at face value, these measurements point to a DE density that rises mildly above its present-day value at intermediate redshifts, with an equation of state that remains marginally phantom at higher redshifts---a preference that persists even in this less degenerate basis, further suggesting it is driven by the data rather than by the choice of parameterization.
While robust across the two parameterizations, however, the overall preference for evolving DE over $\Lambda$CDM is not yet compelling, remaining below the $3\sigma$ level in all of our background-level analyses, with the inclusion of DE perturbations and the full CMB likelihood expected to increase it only marginally (see e.g.~\cite{DESI:2025zgx, DES:2025tir}).
Beyond the statistical considerations discussed in detail in Sec.~\ref{sec:results}, the constraints themselves exhibit features that reinforce the case for caution and warrant further scrutiny.
In particular, our findings reinforce and extend the coincidence identified in Refs.~\cite{Cortes:2024lgw,Efstathiou:2025tie}, where it was shown by pivoting the CPL basis that current data are consistent with the cosmological constant limit $w_p= -1$ at the percent level, despite the overall preference for a substantially varying DE equation of state.
In addition to confirming $w_p\approx -1$ to a precision of $\pm 0.04$ across all data combinations, once SNe are included, the DE density fraction $f_p$ at the pivot is also found to lie only a few percent above unity, consistent with no evolution at the $2\sigma$ level.
This constitutes an additional element of the coincidence, as in the context of an evolving DE scenario there is no a priori reason for the DE density at the pivot---which for $w_p\approx -1$ is expected to peak at this epoch---to coincide so closely with its present-day value.

As a whole, these observations do not settle the question of whether DE evolves, but they illustrate the value of the $(w_p,\,f_p)$ formulation as a complementary diagnostic to the established CPL prescription.
By concentrating the available information into two individually well-measured quantities tied directly to the DE density evolution, it makes the interplay between dataset combinations and the proximity of current constraints to the cosmological constant limit considerably more transparent.
This clarity will only sharpen as next-generation surveys tighten the constraints.

Several extensions to this work merit future investigation.
Our analysis was carried out at the background level, which allowed us to consistently explore regions of parameter space where $f_{\rm DE}$ becomes negative at early times, without the corresponding singularities in $w(z)$ that would compromise the perturbation equations.
However, the constraints obtained here maintain $f_{\rm DE} > 0$ within the 95\% confidence level for all analyses including SNe data, making the extension to include DE perturbations and the full CMB likelihood straightforward, following an approach analogous to that of Ref.~\cite{Cheng:2025lod} for the equivalent pressure parameterization.
This will be particularly interesting in the context of forecasting the constraining power on $(w_p,\,f_p)$ from the combined analysis of next-generation BAO, SNe, and CMB surveys, as recently done in~\cite{LSSTDarkEnergyScience:2026ach} in the context of the CPL parameterization.
Additionally, a dedicated prior sensitivity study of the $(w_p,\,f_p)$ parameterization, analogous to that of Ref.~\cite{Toomey:2025xyo} for CPL, would be valuable to test whether the less degenerate mapping from the DE parameters to the expansion rate demonstrated throughout this work translates into a reduced dependence on prior assumptions, as we expect.
Finally, it will also be important to validate our $(w_p,\,f_p)$ framework against a broader spectrum of evolving DE scenarios beyond thawing quintessence, including effective density evolutions arising from dark matter--dark energy interactions, which have recently attracted considerable interest as viable alternatives to dynamical DE in explaining the deviations from $\Lambda$CDM observed in current data (e.g.~\cite{Li:2026xaz,Khoury:2025txd,Li:2025ula,Paliathanasis:2026ymi}).
We leave the explorations of all of these possibilities to future work.

\begin{acknowledgments}
GM dedicates this work to his newborn daughter, Carla Montefalcone, with love.

We acknowledge the use of \texttt{CLASS}~\cite{Blas:2011rf}, \texttt{GetDist}~\cite{2019arXiv191013970L}, \texttt{MontePython}~\cite{Audren:2012wb, Brinckmann:2018cvx}, and the Python packages \texttt{Matplotlib}~\cite{Hunter:2007mat}, \texttt{NumPy}~\cite{Harris:2020xlr}, and \texttt{SciPy}~\cite{Virtanen:2019joe}.
We also acknowledge the use of the \texttt{MontePython}-based implementations of the DESI-DR2~\cite{Herold:2024nvk} and Pantheon\texttt{+} and DES-Y5 SNe~\cite{Herold:2025hkb} likelihoods, respectively written by Laura Herold and Tanvi Karwal.

We acknowledge the Texas Advanced Computing Center (TACC) at The University of Texas at Austin for providing high-performance computing resources that have contributed to the research results reported within this paper.

GM acknowledges support by the Writing Fellowship of the Graduate School of the College of Natural Sciences at the University of Texas at Austin.
RS acknowledges financial support from STFC Grant No. ST/X508664/1, and the Snell Exhibition of Balliol College, Oxford.

\end{acknowledgments}

\appendix
\section{Analysis of the single-parameter Dark Energy density formulation}\label{app:single_fp}

\begin{figure}[ht!]
    \centering
    \includegraphics[width=\linewidth]{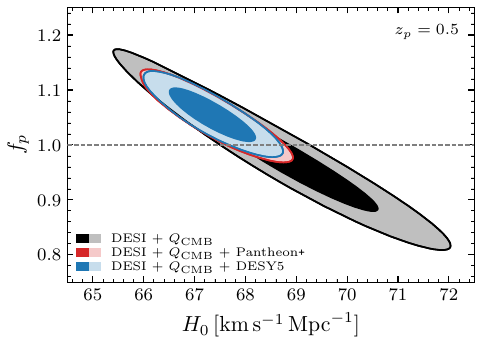}
    \caption{Two-dimensional posteriors in $f_p$ and $H_0$ at the 68\% and 95\% confidence levels from the DESI + $Q_{\rm CMB}$ (black), DESI + $Q_{\rm CMB}$ + Pantheon\texttt{+} (\textcolor{tabred}{red}), and DESI + $Q_{\rm CMB}$ + DESY5 (\textcolor{tabblue}{blue}) dataset combinations, with the pivot redshift set to $z_p=0.5$ as in the main analysis presented in Sec.~\ref{sec:results}.  The dashed \textcolor{tabgray}{gray} line marks the $\Lambda$CDM limit corresponding to $f_p=1$. All three dataset combinations yield constraints consistent with $\Lambda$CDM, with SNe tightening the posteriors significantly while shifting the central value mildly above unity.}
    \label{fig:A_fp}
\end{figure}

\noindent Truncating the polynomial DE density expansion of Eq.~\eqref{eq:fDE_fafb} at first order ($f_b=0$) yields a single-parameter formulation in which the DE density evolves linearly in $(1-a)$. Reparameterizing it in terms of the DE density fraction at the pivot redshift, $f_p\equiv f_{\rm DE}(a_p)$, the normalized DE density reduces to
\begin{equation}
    f_{\rm DE}(z) = 1 + (f_p - 1)\,\frac{z\,(1+z_p)}{z_p\,(1+z)}\,,\label{eq:fDE_fp}
\end{equation}
with the $\Lambda$CDM limit recovered for $f_p=1$. The corresponding equation of state follows from the continuity equation, i.e. Eq.~\eqref{eq:wDE_fafb}, and is fully determined by $f_p$.

We analyze this single-parameter formulation using the same three dataset combinations and methodology described in Sec.~\ref{sec:methodology}, adopting the flat prior $f_p\in[-1,\,3]$ with $z_p=0.5$. Figure~\ref{fig:A_fp} displays the resulting posteriors in the $(f_p,\,H_0)$ plane, while Table~\ref{tab:A1} summarizes the corresponding frequentist significances quantified through the $\Delta\chi^2_{\rm MAP}$ relative to $\Lambda$CDM. For completeness, we also show in Fig.~\ref{fig:A9} the full posterior distributions including the $\Lambda$CDM parameters for all three dataset combinations under consideration. The marginalized $68\%$ C.L. constraints on $f_p$ are:
\begin{equation}
f_p =
\begin{cases}
\,0.99^{+0.05}_{-0.08} & \text{DESI + $Q_{\rm CMB}$}\,,\\[4pt]
\,1.05\pm 0.04 & \text{DESI + $Q_{\rm CMB}$ + Pantheon\texttt{+}}\,,\\[4pt]
\,1.05\pm 0.03 & \text{DESI + $Q_{\rm CMB}$ + DESY5}\,.
\end{cases}
\end{equation}

Without SNe, $f_p$ is consistent with unity and the preference for evolving DE is negligible. The inclusion of SNe shifts the central value mildly above unity and tightens the constraint, but the results remain consistent with $\Lambda$CDM at roughly the $1\sigma$ level across all dataset combinations, with the preference for evolving DE reaching at most $1.7\sigma$ for DESI + $Q_{\rm CMB}$ + DESY5. This preference is weaker by at least $1\sigma$ compared to the two-parameter $(w_p,\,f_p)$ formulation across all dataset combinations (see Table~\ref{tab:2}), indicating that the evidence for evolving DE remains sensitive to the number of free parameters and the specific functional form adopted, thereby reinforcing the case for caution in interpreting the observed marginal deviations from $\Lambda$CDM.

\begin{table}
	\centering
	\begin{tabular}{l c @{\hskip 2em} c}
			\toprule
	Dataset &  $\Delta\chi^2_{\rm MAP}$ & Significance\\
            \midrule[0.065em]
    $\boldsymbol{f_p}{\rm \bf CDM}$ & &\\
    [4pt]
	DESI + $Q_{\rm CMB}$ 	& $-0.18$ & $0.4\,\sigma$	\\
    [4pt]
    DESI + $Q_{\rm CMB}$ + Pantheon\texttt{+} 	& $-2.09$ &	$1.5\,\sigma$\\[4pt]
    DESI + $Q_{\rm CMB}$ + DESY5	& $-2.96$	 & $1.7\,\sigma$\\[2pt]
	\bottomrule
	\end{tabular}
	\caption{Same as Table~\ref{tab:2} for the single-parameter $f_p$CDM formulation. Since $\Lambda$CDM is nested within this model as $f_p=1$, the significance is computed assuming one extra free parameter. The preference for evolving DE remains below $2\sigma$ across all dataset combinations.}
	\label{tab:A1}
\end{table}

\begin{figure*}[ht!]
    \centering
    \includegraphics[width=0.9\linewidth]{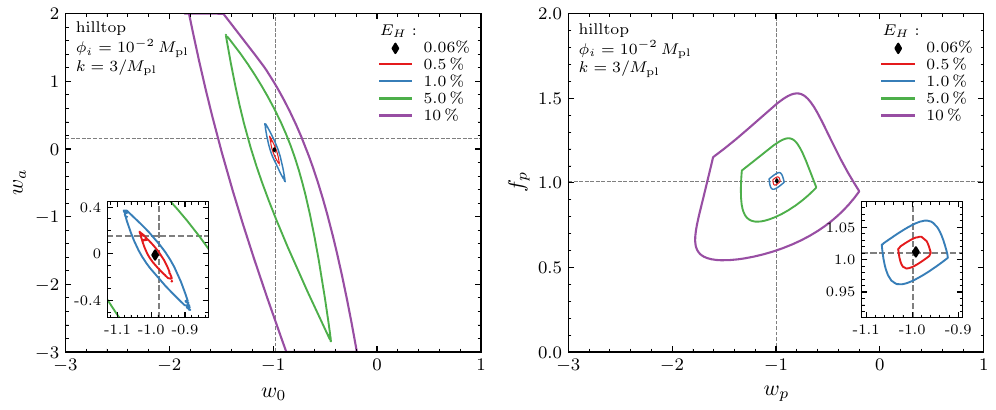}
    \caption{Same as Fig.~\ref{fig:1} for a hilltop quintessence, Eq.~\eqref{eq:hill}, with $k=3/M_{\rm pl}$ and $\phi_i=10^{-2}\,M_{\rm pl}$. The same qualitative features are reproduced, with the $(w_p,\,f_p)$ plane again exhibiting compact contours that recover the physical DE properties at the pivot.}
    \label{fig:A1}
\end{figure*}

Nevertheless, as anticipated in Sec.~\ref{sec:pivot}, the weakness of this preference should not be taken entirely at face value. As pointed out in Ref.~\cite{Linder:2007ka} in the context of the constant DE equation of state extension $w$CDM, a single DE parameter must simultaneously accommodate the tightly constrained CMB distance to last scattering and the lower-redshift BAO and SNe measurements, leaving little freedom to deviate from $\Lambda$CDM in either regime without spoiling the fit in the other. Moreover, if the actual signal in the data is a time-varying deviation from $\Lambda$CDM---as suggested by the two-parameter analyses in the main text---a single degree of freedom will inevitably dilute it by enforcing a rigid functional shape across the full redshift range. These limitations motivate the two-parameter $(w_p,\,f_p)$ formulation developed in the main text, which provides the additional freedom needed to accommodate the possible rich redshift evolution of an evolving DE component~(see also Fig.~\ref{fig:A5} in Appendix~\ref{app:validation}).

\section{Supplementary Validation Tests}\label{app:validation}

\noindent In this appendix we supplement the exponential quintessence benchmark of Sec.~\ref{sec:validation} with an analogous analysis for a hilltop potential, and further validate the $(w_p,\,f_p)$ formulation through mock data analyses that test parameter recovery and the fidelity of the reconstructed DE evolution against known fiducial cosmologies.
\subsection{Hilltop quintessence benchmark}\label{app:hilltop}

\noindent Figure~\ref{fig:A1} shows the contours of the maximum relative error $E_H$, Eq.~\eqref{eq:EH}, in the $(w_p,\,f_p)$ and $(w_0,\,w_a)$ planes for a hilltop quintessence potential, Eq.~\eqref{eq:hill}, with steepness parameter $k=3/M_{\rm pl}$ and initial value of the field $\phi_i=10^{-2}\,M_{\rm pl}$. This illustration complements the exponential case discussed in Sec.~\ref{sec:validation}, confirming that the advantages of the density-level parameterization are not specific to a particular potential but hold across representative quintessence scenarios. 

Here, both parameterizations achieve a sub-percent best-fit accuracy with $E_H\approx 0.06\%$, slightly better than the exponential case. The structure of the error contours between the $(w_p,\,f_p)$ and $(w_0,\,w_a)$ formulations is again markedly different and reproduces the same qualitative features identified in Fig.~\ref{fig:1}. The $(w_0,\,w_a)$ plane displays elongated contours along the $\Delta w_a/\Delta w_0\approx -5$ degeneracy direction, with the Taylor coefficients of $w(a)$
at $a=1$ falling outside the 1\% error contour, while the $(w_p,\,f_p)$ plane exhibits compact, non-degenerate contours where the physical values of the DE density fraction and equation of state at the pivot nearly coincide with the best-fit point.

\begin{figure}[hb!]
    \centering
    \includegraphics[width=\linewidth]{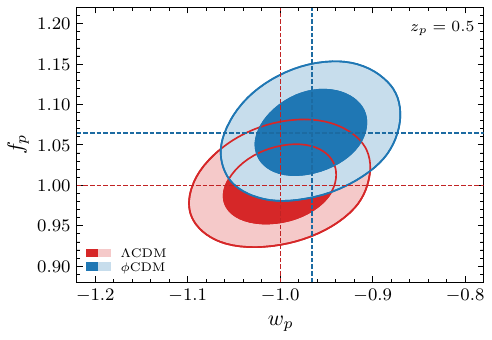}
    \caption{Two-dimensional posteriors in $w_p$ and $f_p$ at the 68\% and 95\% confidence levels from the analysis of mock DESI + $Q_{\rm CMB}$ + DESY5 data generated from a $\Lambda$CDM (\textcolor{tabred}{red}) and an exponential quintessence $\phi$CDM (\textcolor{tabblue}{blue}) fiducial cosmology, as described in the text. The pivot redshift is set to $z_p=0.5$ as done for the main analyses in Sec.~\ref{sec:results}. Dashed lines in the corresponding colors mark the fiducial $(w_p,\,f_p)$ values for each scenario, both of which fall well within the 68\% confidence region, confirming unbiased parameter recovery.}
    \label{fig:A3}
\end{figure}

\begin{figure*}
    \centering
    \includegraphics[width=0.9\linewidth]{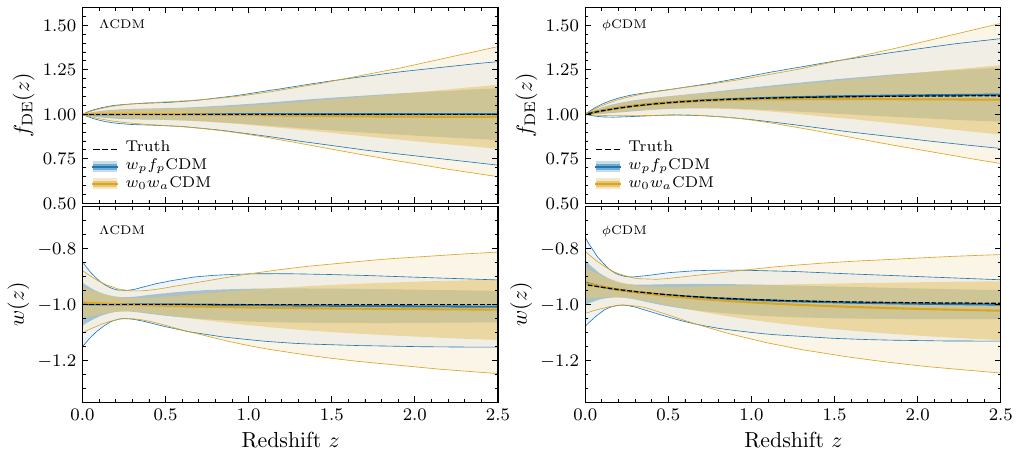}
    \caption{{\it Left panels:} Normalized dark energy density, $f_{\rm DE}(z) \equiv \rho_{\rm DE}(z)/\rho_{\rm DE,0}$ ({\it top}), and corresponding equation of state parameter, $w(z)$ ({\it bottom}), as a function of redshift from the analysis of mock DESI + $Q_{\rm CMB}$ + DESY5 data generated from a fiducial $\Lambda$CDM cosmology. Results are shown for the $(w_p,\,f_p)$ (\textcolor{tabblue}{blue}) and $(w_0,\,w_a)$ (\textcolor{goldenrod}{gold}) parameterizations, with solid lines indicating the posterior median and increasingly lighter shading denoting the 68\% and 95\% confidence level bands. The dashed black line marks the fiducial $\Lambda$CDM evolution, namely $f_{\rm DE}=1$ and $w=-1$. {\it Right panels:} Same as the {\it Left panels} but for a fiducial exponential quintessence ($\phi$CDM) cosmology with $\lambda=0.7$, with the dashed black line showing the true quintessence evolution. Both parameterizations recover the true underlying evolution within the expected confidence bands, with the $(w_p,\,f_p)$ formulation yielding tighter reconstruction bands at $z\gtrsim 2$, as discussed in the text. }
    \label{fig:A4}
\end{figure*}

\subsection{Mock data analysis}\label{app:mocks}

\noindent To further validate the $(w_p,\,f_p)$ formulation, we perform a set of mock data analyses designed to test parameter recovery and the fidelity of the reconstructed DE evolution against known fiducial cosmologies. 

Each mock dataset comprises simulated DESI DR2 BAO measurements, DESY5 SNe distance moduli, and the compressed $Q_{\rm CMB}$ likelihood, matching the data combination used in our main analysis. The BAO mock data consist of the theory predictions for the comoving distance $D_M$, the Hubble distance $D_H$, and the spherically-averaged distance $D_V\equiv (zD_M^2 D_H)^{1/3}$, all measured relative to the sound horizon $r_d$ at the DESI DR2 redshift bins. These are adopted as the mean data vector, together with the actual DESI DR2 covariance matrix. Similarly, the SNe mock data are constructed by evaluating the theory distance moduli at the observed DESY5 redshifts, using the DESY5 covariance matrix to define the uncertainties and correlations. This noiseless approach isolates the performance of the parameterization itself from statistical fluctuations in any specific realization of the data.

We consider in particular two fiducial cosmologies. The first assumes $\Lambda$CDM with cosmological parameters set to their {\it Planck} best-fit values~\cite{Planck:2018vyg}. The second is the exponential quintessence scenario of Eq.~\eqref{eq:exp} with $\lambda=0.7$, the same benchmark used in Sec.~\ref{sec:validation}, which we refer to as $\phi$CDM. Although these two realizations do not exhaust the space of possible DE models, they bracket qualitatively distinct physical behaviors---a static cosmological constant and a thawing scalar field---and provide a meaningful test of whether our polynomial DE density formulation introduces systematic biases in parameter recovery or in the reconstructed DE evolution.

We analyze both mock datasets using the same MCMC methodology, priors, and pipeline described in Sec.~\ref{sec:methodology}, analyzing the $(w_p,\,f_p)$ and CPL DE parameterizations alongside $\Lambda$CDM to each mock in turn.

Figure~\ref{fig:A3} shows the corresponding two-dimensional posteriors in the $(w_p,\,f_p)$ plane for both mock datasets. In the $\Lambda$CDM case, the fiducial values $(w_p,\,f_p)=(-1,\,1)$ are recovered well within the 68\% confidence region. Similarly, for the $\phi$CDM mock, the fiducial $(w_p,\,f_p)$ values---determined by the quintessence dynamics at the pivot redshift---fall well within the expected confidence regions, confirming that the parameterization does not systematically bias the constraints away from the true underlying cosmology in either scenario.

Figure~\ref{fig:A4} corroborates these findings by displaying the reconstructed evolution of the normalized DE density $f_{\rm DE}(z)$ and equation of state $w(z)$ for both mock datasets, showing that the true underlying evolution is recovered within the expected confidence bands across the relevant redshift range. 

The figure also overlays the corresponding CPL reconstructions for comparison. The two parameterizations yield consistent median reconstructions, but at higher redshifts, $z\gtrsim 2$, the CPL uncertainty bands become visibly broader than those of the $(w_p,\,f_p)$ formulation. As discussed in Sec.~\ref{sec:geometric}, this is a consequence of the nonlinear exponential mapping from $(w_0,\,w_a)$
to $f_{\rm DE}$, Eq.~\eqref{eq:fDE}, which amplifies parameter uncertainties when propagated to the DE density, particularly at higher redshifts. The polynomial density formulation avoids this amplification by construction, as parameter uncertainties simply propagate as polynomial corrections to $f_{\rm DE}$ rather than through an exponential, resulting in a more controlled growth of the reconstruction uncertainty with redshift. This advantage, while effectively negligible for the particular DE evolution preferred by current data as noted in Sec.~\ref{sec:results}, becomes more apparent in the $\phi$CDM mock, where the monotonically evolving DE density probes the regime in which the two mappings differ most.

A similar pattern is visible in the reconstructed equation of state $w(z)$, where the $(w_p,\,f_p)$ formulation also yields tighter constraints at higher redshifts. This is expected, as in our formulation $w(z)$ is derived directly from the polynomial density evolution through the continuity equation, Eq.~\eqref{eq:wDE_fafb}, inheriting its controlled uncertainty growth, whereas in CPL the relatively broad $w_a$ posterior drives the reconstructed $w(z)$ to increasingly uncertain values at earlier times.

\begin{figure}
    \centering
    \includegraphics[width=\linewidth]{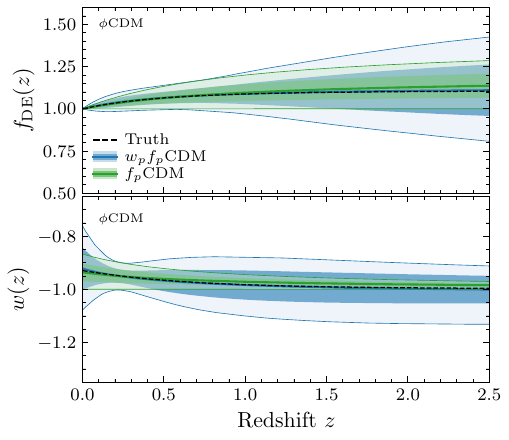}
    \caption{Same as Fig.~\ref{fig:A4} for the $\phi$CDM fiducial cosmology in particular, comparing the two-parameter $(w_p,\,f_p)$ (\textcolor{tabblue}{blue}) and single-parameter $f_p$ (\textcolor{tabgreen}{green}) reconstructions. The dashed black line shows the true quintessence evolution. While both formulations recover the fiducial evolution within the 68\% C.L. bands, the $f_p$-only reconstruction displays a mild but consistent tendency to overestimate $f_{\rm DE}$ and underestimate $w$ at higher redshifts, reflecting the limited flexibility of a single degree of freedom to capture the curvature of a benchmark DE density evolution.}
    \label{fig:A5}
\end{figure}

Finally, while not shown here, we also validated the single-parameter $f_p$ formulation of Appendix~\ref{app:single_fp} against the same mock datasets, finding an accurate recovery of the true DE evolution within the confidence bands for both fiducial cosmologies. However, in the $\phi$CDM scenario, minor systematic differences with respect to the two-parameter $(w_p,\,f_p)$ reconstruction are already discernible. As shown in Fig.~\ref{fig:A5}, which compares the reconstructed $f_{\rm DE}(z)$ and $w(z)$ from the two formulations, the single-parameter model mildly overestimates $f_{\rm DE}$ while underestimating $w$ at higher redshifts. Although these deviations remain well within the 68\% C.L. bands, they display a consistent pattern reflecting the structural rigidity discussed in Appendix~\ref{app:single_fp}, where a single degree of freedom enforces a fixed functional shape that cannot fully accommodate the curvature of the true DE density evolution.

\section{Supplementary parameter constraints and reconstructions}\label{app:posteriors}

\begin{figure}[ht!]
    \centering
   \includegraphics[width=\linewidth]{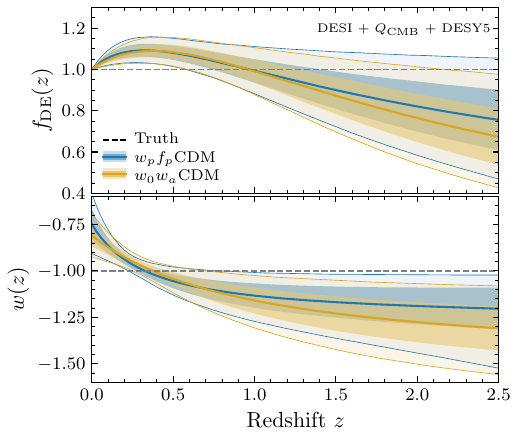}
    \caption{Same as Fig.~\ref{fig:3} for the DESI + $Q_{\rm CMB}$ + DESY5 dataset combination. The reconstructions are consistent with the DESI + $Q_{\rm CMB}$ + Pantheon\texttt{+} analysis, with both the departure of $f_{\rm DE}$ from unity and the phantom preference in $w(z)$ mildly more pronounced. }
    \label{fig:A2}
\end{figure}

\noindent In this appendix we collect supplementary posterior distributions and DE reconstructions that complement the main results of Sec.~\ref{sec:results} and the single-parameter analysis of Appendix~\ref{app:single_fp}. 

\begin{figure*}
    \centering
    \includegraphics[width=0.95\linewidth]{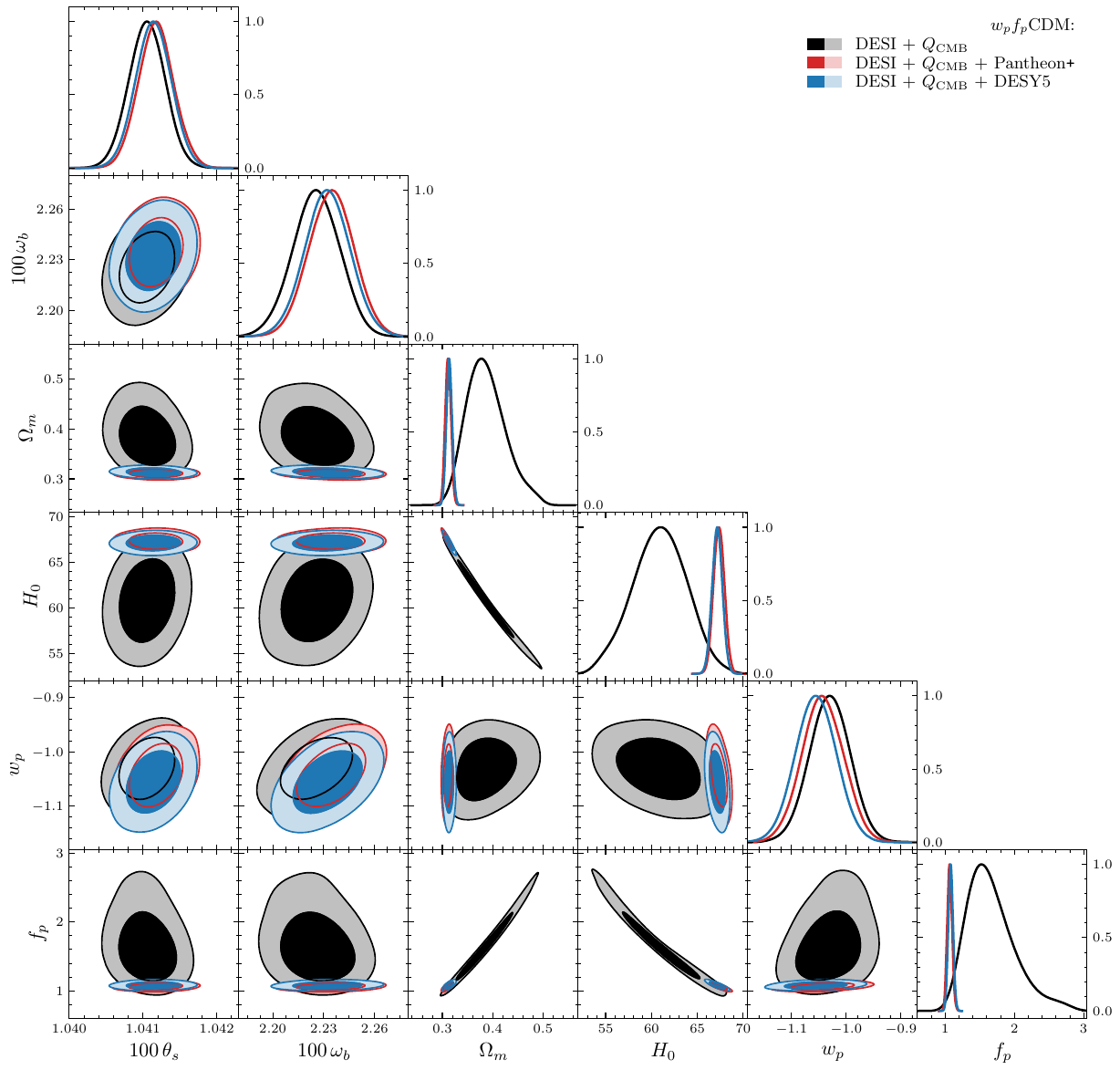}
    \caption{Marginalized two-dimensional posterior distributions for the $\Lambda$CDM parameters ($100\,\theta_s$, $100\,\omega_b$, $\Omega_{\rm m}$, $H_0$) and the DE parameters ($w_p$, $f_p$) from the DESI + $Q_{\rm CMB}$ (black), DESI + $Q_{\rm CMB}$ + Pantheon\texttt{+} (\textcolor{tabred}{red}), and DESI + $Q_{\rm CMB}$ + DESY5 (\textcolor{tabblue}{blue}) dataset combinations in the $w_pf_p$CDM analysis. Contours correspond to the 68\% and 95\% confidence levels, with the one-dimensional marginalized posteriors for each parameter displayed along the diagonal. SNe breaks the degeneracy between $f_p$ and $H_0$ present in the DESI + $Q_{\rm CMB}$ analysis, tightening all constraints and shifting them toward their $\Lambda$CDM values.}
    \label{fig:A6}
\end{figure*}

\begin{figure*}
    \centering
    \includegraphics[width=0.9\linewidth]{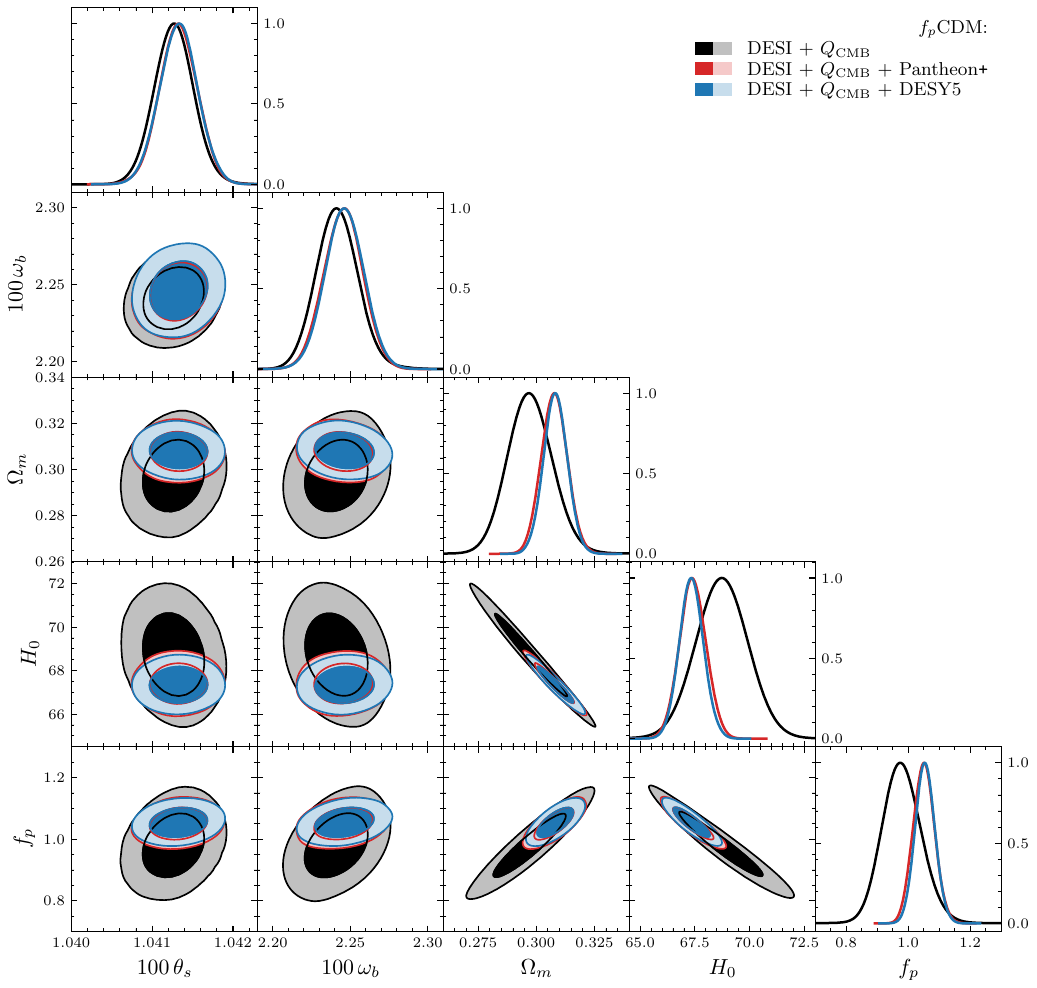}
    \caption{Same as Fig.~\ref{fig:A6} for the single-parameter $f_p$CDM analysis of Appendix~\ref{app:single_fp}, showing $100\,\theta_s$, $100\,\omega_b$, $\Omega_{\rm m}$, $H_0$, and $f_p$. The posteriors are considerably tighter than in the two-parameter case, with all three dataset combinations yielding constraints consistent with the $\Lambda$CDM limit $f_p=1$.}
    \label{fig:A7}
\end{figure*}

\begin{figure*}
    \centering
    \includegraphics[width=0.8\linewidth]{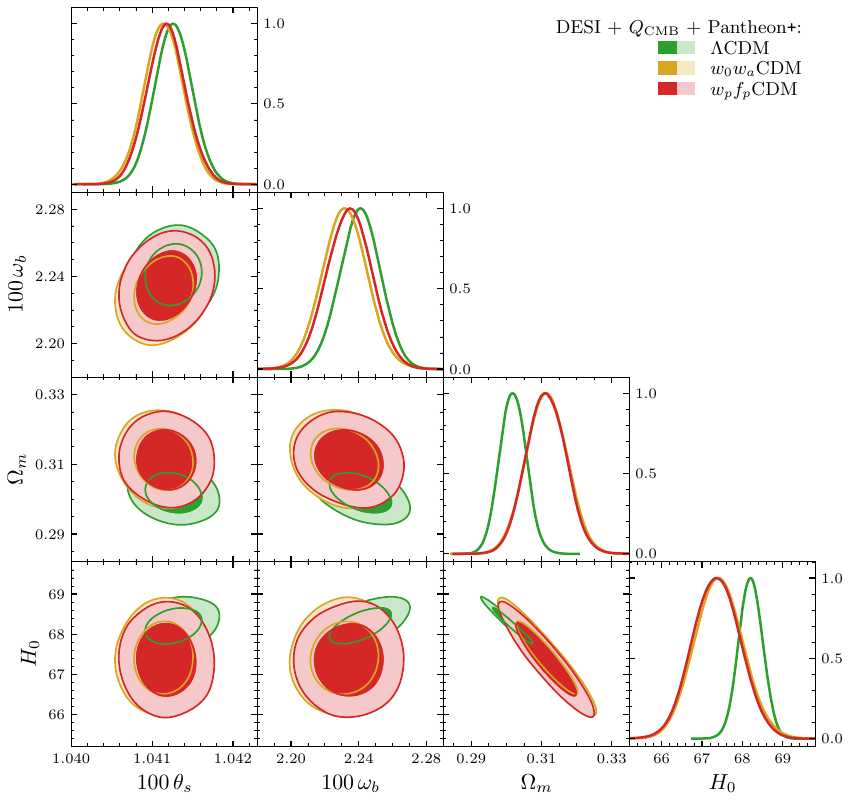}
    \caption{Marginalized two-dimensional posterior distributions for $100\,\theta_s$, $100\,\omega_b$, $\Omega_{\rm m}$, and $H_0$ from the DESI + $Q_{\rm CMB}$ + Pantheon\texttt{+} dataset combination, comparing the $\Lambda$CDM (\textcolor{tabgreen}{green}), $w_0w_a$CDM (\textcolor{goldenrod}{gold}), and $w_pf_p$CDM (\textcolor{tabred}{red}) analyses. Contours correspond to the 68\% and 95\% confidence levels, with the one-dimensional marginalized posteriors for each parameter displayed along the diagonal. The $w_0w_a$CDM and $w_pf_p$CDM analyses yield tightly consistent constraints on all parameters, with $\Omega_{\rm m}$ and $H_0$ systematically shifted to mildly higher and lower values, respectively, relative to the $\Lambda$CDM analysis.}
    \label{fig:A8}
\end{figure*}

\begin{figure*}
    \centering
    \includegraphics[width=0.8\linewidth]{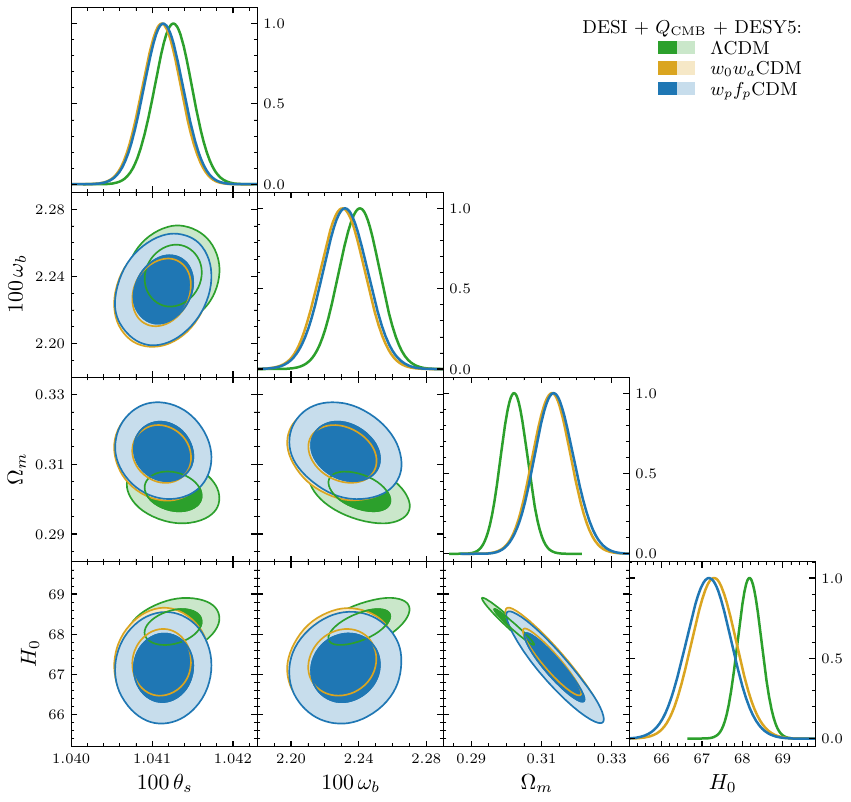}
    \caption{Same as Fig.~\ref{fig:A8} for the DESI + $Q_{\rm CMB}$ + DESY5 dataset combination, with the $w_pf_p$CDM analysis now shown in \textcolor{tabblue}{blue}. The same pattern of consistent constraints between the $w_0w_a$CDM and $w_pf_p$CDM analyses is observed, with both SNe samples yielding compatible results.}
    \label{fig:A9}
\end{figure*}

Figure~\ref{fig:A2} displays the reconstructed $f_{\rm DE}(z)$ and $w(z)$ from the DESI + $Q_{\rm CMB}$ + DESY5 dataset combination, the counterpart of Fig.~\ref{fig:3} in the main text which uses Pantheon\texttt{+}. Figures~\ref{fig:A6} and~\ref{fig:A7} show the full posterior distributions of the key cosmological parameters in the $(w_p,\,f_p)$ and single-parameter $f_p$ analyses, respectively, for all three dataset combinations. Finally, Figures~\ref{fig:A8} and~\ref{fig:A9} compare the constraints on the $\Lambda$CDM parameters under the $\Lambda$CDM, $w_0w_a$CDM, and $w_pf_p$CDM analyses, for the DESI + $Q_{\rm CMB}$ + Pantheon\texttt{+} and the DESI + $Q_{\rm CMB}$ + DESY5 data combinations, respectively. 
Of particular note in these figures is the strong degeneracy between $f_p$ and $H_0$ present in the DESI + $Q_{\rm CMB}$ analysis, which is effectively broken by the independent $\Omega_{\rm m}$ constraint that SNe provide, as discussed in Sec.~\ref{sec:results}. With SNe included, the $w_0w_a$CDM and $w_pf_p$CDM analyses yield tightly consistent constraints among one another on all $\Lambda$CDM parameters, with $\Omega_{\rm m}$ and $H_0$ systematically shifted to mildly higher and lower values, respectively, relative to their $\Lambda$CDM counterparts---a pattern that underlies the marginal preference for evolving DE reported in Sec.~\ref{sec:results}.

\onecolumngrid

\clearpage
\twocolumngrid
\bibliographystyle{apsrev4-1}
\bibliography{bibl}

\end{document}